\documentclass[12pt,a4paper]{article}
\usepackage{amssymb,amsmath}
\usepackage{epsfig}
\usepackage{times}

\renewcommand{\theequation}{\arabic{section}.\arabic{equation}}

\newcommand{\Z}{\mathbb{Z}}
\newcommand{\R}{\mathbb{R}}
\newcommand{\C}{\mathbb{C}}
\newcommand{\N}{\mathbb{N}}
\newcommand{\E}{\mathbb{E}}
\newcommand{\Q}{\mathbb{Q}}
\newcommand{\EE}{\mathsf{E}}
\renewcommand{\P}{\mathbb{P}}
\newcommand{\supp}{{\ensuremath{\rm supp}}}
\newcommand{\tr}{\mathrm{tr}}
\newcommand{\const}{\mathrm{const}}
\newcommand{\sign}{\mbox{\rm sign}}
\newcommand{\SL}{\mbox{\rm SL}}
\newcommand{\GL}{\mbox{\rm GL}}
\renewcommand{\Im}{\ensuremath{\rm Im}}
\renewcommand{\Re}{\ensuremath{\rm Re}}
\title{Scattering Theory Approach \\ to Random Schr\"odinger Operators\\
in One Dimension}
\author{V. Kostrykin\thanks{e-mail: kostrykin@ilt.fhg.de}\\
Lehrstuhl f\"ur Lasertechnik\\ Rheinisch
- Westf\"alische Technische Hochschule Aachen\\
Steinbachstr. 15, D-52074 Aachen, Germany
\and
and\\
 R. Schrader\thanks{e-mail: schrader@physik.fu-berlin.de}\\
 Institut f\"{u}r Theoretische Physik\\ Freie Universit\"{a}t Berlin, Arnimallee
14\\ D-14195 Berlin, Germany}
\date{February 19,1998}

\begin{document}
\maketitle

\footnotetext[1]{Published in Reviews in Marthematical Physics \textbf{11}, 187 -- 242 (1999)}

\begin{abstract}
Methods from scattering theory are introduced to analyze random Schr\"o\-din\-ger
operators in one dimension by applying a volume cutoff to the potential. The key
ingredient is the Lifshitz-Krein spectral shift function, which is related to the
scattering phase by the theorem of Birman and Krein. The spectral shift density is defined
as the ``thermodynamic limit" of the spectral shift function per unit length of the
interaction region. This density is shown to be equal to the difference of the densities
of states for the free and the interacting Hamiltonians. Based on this construction, we
give a new proof of the Thouless formula. We provide a prescription how to obtain the
Lyapunov exponent from the scattering matrix, which suggest a way how to extend this
notion to the higher dimensional case. This prescription also allows a characterization of
those energies which have vanishing Lyapunov exponent.
\end{abstract}

\pagebreak

\section{Introduction}
\markright{1 Introduction}

In this paper we consider random Schr\"{o}dinger operators $H(\omega)$
in $L^2(\R)$ of the form
\begin{equation}\label{1.1}
H(\omega)=H_0+\sum_{j\in{\Z}}
\alpha_j(\omega)f(\cdot-j),\ H_0=-\frac{d^2}{dx^2},
\end{equation}
where $\{\alpha_j(\omega)\}_{j\in{\Z}}$ is a sequence of i.i.d. (independent, identically
distributed) variables having a common density $\varphi$ (i.e. $\P\{\alpha_j\in
dy\}=\varphi(y)dy$), which is continuous and has support in the finite interval
$[\alpha_-,\alpha_+]$. In what follows we always suppose that the single-site potential
$f$ is piecewise continuous with $\supp f\subseteq [-1/2,1/2]$. Moreover, we require that
$f\geq 0$. The spectral properties of (\ref{1.1}) were studied in detail in
\cite{KiMa,Delyon,KiKoSi}. The results are most complete for the case when $f$ is the
point interaction (see \cite{Albeverio:book}).

The main idea of our approach is to approximate the operator
$H(\omega)$ (\ref{1.1}) by means of the sequence
\begin{equation}\label{1.2}
H^{(n)}(\omega)=H_0+\sum_{j=-n}^{j=n}
\alpha_j(\omega)f(\cdot-j)
\end{equation}
with unchanged $H_0$, which converges to $H(\omega)$ in the strong resolvent sense. This
approximation gives the opportunity to use scattering theory in order to study the
spectral properties of the limiting operator (\ref{1.1}). In fact, we show how to recover
the spectral characteristics of $H(\omega)$ from the limiting behaviour of the spectral
characteristics of $H^{(n)}(\omega)$ in the ``large support" limit $n\rightarrow\infty$.

One of the important ingredients of our approach is the Lifshitz-Krein spectral shift
function (see \cite{BiYa} for a review). Recently it has received renewed interest due to
its applications to different problems in the theory of Schr\"odinger operators
\cite{Jensen:Kato,Schrader,Guillope,Bolle,BMS,Jensen,Robert,KoSch1,GKS,Bruneau,Gesztesy:Simon,Simon:preprint,KoSch2,Mueller}.
In the context of our approach the spectral shift function naturally replaces the
eigenvalue counting function, which is usually used to construct the density of states for
the operator (\ref{1.1}). The celebrated Birman-Krein theorem \cite{BiKr} relates the
spectral shift function to scattering theory. In fact, up to a factor $-\pi^{-1}$ it may
be identified with the scattering phase for the pair ($H^{(n)}(\omega)$, $H_0$), i.e.
$\xi^{(n)}(E;\omega)=-\pi^{-1}\delta^{(n)}(E;\omega)$ when $E>0$,
\begin{displaymath}
\delta^{(n)}(E;\omega)=\frac{1}{2i}\log\det S^{(n)}(E;\omega)=
\frac{1}{2i}\log\det \left(\begin{array}{lr}
                  T_\omega^{(n)}(E) & R_\omega^{(n)}(E) \\
                  L_\omega^{(n)}(E) & T_\omega^{(n)}(E)
                  \end{array}\right).
\end{displaymath}
Here $|T^{(n)}(E)|^2$ and $|R^{(n)}(E)|^2=|L^{(n)}(E)|^2$ have the meaning of transmission
and reflection coefficients, respectively, such that $|T^{(n)}(E)|^2+|R^{(n)}(E)|^2=1$.
For $E<0$\hspace{1mm} $\xi^{(n)}(E;\omega)$ equals minus the counting function of
$H^{(n)}(\omega)$.

These two properties of the spectral shift function, namely its relation to scattering
theory and its replacement of the counting function in the presence of an absolutely
continuous spectrum convinced the authors already some time ago that the spectral shift
function could be applied to the theory of random Schr\"odinger operators. In fact, our
previous articles \cite{KoSch1,GKS,KoSch2} were in part preparatory investigations aiming
at such an application. In \cite{GKS} we proved convexity and subadditivity properties of
the spectral shift function with respect to the potential and the coupling constant,
respectively. Such properties often show up when considering thermodynamic limits in
statistical mechanics. In \cite{KoSch1,KoSch2} we studied cluster properties when the
potential is a sum of two terms and the center of one is moved to infinity. Again such
properties play an important role in statistical mechanics as well as in quantum field
theory. Possible applications of the theory of the spectral shift function to random
Schr\"odinger operators have also been envisaged by Simon \cite{Simon:preprint}. Some
other applications of the scattering theory in one dimension to the study of spectral
properties of Schr\"odinger operators with periodic or random potentials can be found in
\cite{Keller,Rorres} and \cite{KiKoSi} respectively.

Below we prove some new inequalities for the spectral shift function, which reflect its
``additivity'' properties with respect to the potential being the sum of two terms with
disjoint supports. These inequalities are closely related to the Aktosun factorization
formula (\ref{Aktosun.form}) \cite{Aktosun} (see also
\cite{Bianchi,Aktosun:Klaus:Mee,Bianchi2}). Combined with the superadditive ergodic
theorem they will allow us to prove the almost sure existence of the limit
\begin{equation}\label{xi}
\xi(E)=\lim_{n\rightarrow\infty}\frac{\xi^{(n)}(E;\omega)}{2n+1},
\end{equation}
which we call the spectral shift density. We prove the equality $\xi(E)=N_0(E)-N(E)$,
where $N(E)$ and $N_0(E)=\pi^{-1}[\max(0,E)]^{1/2}$ are the integrated density of states
of the Hamiltonians $H(\omega)$ and $H_0$, respectively. Also we reconsider the Aktosun
factorization formula and show that it is a direct consequence of the propagator property
of the fundamental (or transfer) matrix of the Schr\"odinger equation.

Another very important quantity associated with the Hamiltonian (\ref{1.1}) is the
Lyapunov exponent $\gamma(E)$. In particular, according to the Ishii-Pastur-Kotani theorem
\cite{Kotani} the set $\{E:\gamma(E)=0\}$ is the essential support of the absolute
continuous part of the spectral measure for $H(\omega)$ (\ref{1.1}).

We will establish that for $E>0$ the function
$-\gamma(E)+i\pi(N(E)-N_0(E))$ can be interpreted as the density of the logarithm of the
transmission amplitude $T(E)$,
\begin{equation}\label{1.3}
\lim_{n\rightarrow\infty}\frac{\log T_\omega^{(n)}(E)}{2n+1}=
-\gamma(E)-i\pi\xi(E).
\end{equation}
The connection between the Lyapunov exponent and the transmission coefficient
$|T_\omega^{(n)}(E)|$ was recognized long ago
\cite{Lifshitz:Gredeskul:Pasur,Lifshitz:book}, though a rigorous analysis was still
absent. It is well known (see e.g. \cite{CaLa}) that the function $w(E)=-\gamma(E)+i\pi
N(E)$ can be analytically continued in the complex half-plane ${\Im} E>0$ as a Nevanlinna
function (i.e. an analytic function which maps the upper complex half plane into itself).
We will recover this property of $w(E)$ directly from the analytic properties of the
transmission amplitudes $T_\omega^{(n)}(E)$. Moreover, as a direct consequence of these
properties we give a new proof of the Thouless formula
\begin{equation}\label{Thouless}
\gamma(E)-\gamma_0(E)=-\int_{\R}\log|E-E'|d\xi(E'),
\end{equation}
where $\gamma_0(E)=[\max(0,-E)]^{1/2}$ is the Lyapunov exponent
for $H_0$.

Also we prove a new representation for the Lyapunov exponent for positive energies,
\begin{equation}\label{xxx}
\gamma(E)=\lim_{n\rightarrow\infty}\frac{1}{2n+1}\log\left\|\prod_{j=-n}^n
\widetilde{\Lambda}_{\alpha_j(\omega)}(E) \right\|,
\end{equation}
where
\begin{equation}\label{L.tilde}
\widetilde{\Lambda}_{\alpha}(E)=\left(\begin{array}{cc}
                                 \frac{e^{-i\sqrt{E}}}{T_\alpha(E)} &
                                 -\frac{R_\alpha(E)}{T_\alpha(E)}\\
                                 \frac{L_\alpha(E)}{T_\alpha(E)} &
                                 \frac{e^{i\sqrt{E}}}{T_\alpha(E)^\ast}
                                 \end{array}\right),
\end{equation}
and $T_\alpha(E)$, $R_\alpha(E)$, $L_\alpha(E)$ are elements of the S-matrix at energy $E$
for the pair of Hamiltonians ($H_0+\alpha f$, $H_0$). This representation allows us to
apply the theory of random matrices to prove that $\gamma(E)>0$ for a.e. $E>0$ almost
surely, which in turn by Ishii-Pastur-Kotani theorem implies that the spectrum of $H$ has
no absolute continuous component in $(0,\infty)$. We give also an explicit description of
the set of special energies where $\gamma(E)=0$. To our best knowledge this set was known
explicitly only for the two particular cases when the single-site potential $f$ is a
$\delta$-potential or the characteristic function of the interval $[-1/2,1/2]$ (see
\cite{Figotin:Pastur}).

We express the density of states $N(E)$ for positive energies in terms of the product of
matrices (\ref{L.tilde}),
\begin{equation}\label{1.6}
N(E)=\mp\frac{1}{\pi}\lim_{n\rightarrow\infty}\frac{1}{2n+1}
\arg\left(e_\pm, \prod_{j=-n}^n \widetilde{\Lambda}_{\alpha_j(\omega)}(E) e_\pm\right),
\end{equation}
with $e_+=(1,0)^T$, $e_-=(0,1)^T$ and $(\cdot,\cdot)$ being the inner product in $\C^2$.
This representation is similar to the definition of the density of states through the
rotation number of the fundamental solution of the Schr\"odinger equation
\cite{Johnson:Moser}.

Formulae (\ref{xi}), (\ref{1.3}), (\ref{xxx}), and (\ref{1.6}) also provide very simple
and efficient numerical algorithms to compute the density of states and the Lyapunov
exponent of disordered systems in one dimension. In this context we also remark that the
representations (\ref{xi}) and (\ref{1.6}) for finite $n$ and $E>0$ give smooth
approximations to the spectral shift density and density of states, respectively. This
contrasts with the usual procedure (see e.g. \cite{Figotin:Pastur}), where the density of
states is approximated by step-like functions. As an example in Section 4 we have
calculated the spectral shift density for the deterministic Kronig-Penney model.

Our approach is also applicable to deterministic Hamiltonians of the form $H=H_0+V$ with a
potential $V$, which is supposed to be uniformly in $L^1_\mathrm{loc}(\R)$ and for
technical reasons also bounded, but without any assumption on its decay at infinity. Let
$\{y_j\}_{j\in\Z}$ be a sequence of real numbers such that $y_j\rightarrow\pm\infty$ as
$j\rightarrow\pm\infty$ and $y_j<y_{j+1}$ for all $j\in\Z$. In this case we approximate
$H$ through $H^{(n)}=H_0+V\chi^{(n)}$ with $\chi^{(n)}$ being the characteristic function
of the interval $[y_{-n},y_n]$. Again we show that the relations analogous to (\ref{xi}),
(\ref{1.3}), (\ref{xxx}), (\ref{1.6}) hold.

An extension of ideas developed in the present paper to the case of higher dimensions will
be given in \cite{KoSch3}.

\textbf{Acknowledgements:} We are indebted to J.M. Combes and V. Enss for valuable remarks.

\section{Auxiliary Results}
\setcounter{equation}{0}
\markright{2 Auxiliary Results}

We start with a short discussion of some important properties of the spectral shift
function $\xi(E)=\xi(E;H,H_0)$ for a pair of self-adjoint Hamiltonians $H=H_0+V$ and
$H_0=-d^2/dx^2$ on $L^2(\R)$ with domain of definition  being the Sobolev space
$W^{2,2}(\R)$ (see e.g. \cite{RS2} for the definition). Here $V$ denotes the
multiplication operator by the real valued function $V(x)$, which is supposed to satisfy
\begin{equation}\label{condition}
\int_\R (1+|x|^2)|V(x)|dx <\infty.
\end{equation}
The spectral shift function is defined by the trace formula
\begin{equation}\label{one}
\tr\left(\phi(H)-\phi(H_0)\right)=\int_\R\phi'(E)\xi(E)dE,
\end{equation}
which is valid for a wide class of functions $\phi$ (see \cite{BiYa}). For instance,
elements in $C_0^\infty(\R)$ are in this class. Relation (\ref{one}) defines $\xi(E)$ only
up to an additive constant. We fix this ambiguity by the condition that $\xi(E)=0$ for all
$E$ below the spectrum $\sigma(H)$ of $H$.

With this normalization condition the function $\xi(E)$ for $E<0$ equals minus the number
of the eigenvalues of $H=H_0+V$ less than $E$,
\begin{equation}\label{two}
\xi(E-0)=-n_{E-0}(V).
\end{equation}
For $E>0$ the relation between $\xi(E)$ and the scattering matrix $S(E)$ is given by
the celebrated Birman-Krein theorem \cite{BiYa},
\begin{equation}\label{three}
\log\det S(E)=-2\pi i\xi(E).
\end{equation}
Here the branch of the logarithm is fixed by the condition $\xi(E)\rightarrow 0$ as
$E\rightarrow\infty$ (see Lemma 2.1 below). Since $S(E)$ is continuous for all $E>0$ \cite{Faddeev}, so is
$\xi(E)$.

The relations (\ref{two}) and (\ref{three}) define $\xi(E)$ everywhere, i.e. also for
$E<0,$ except at the finite number of points of discontinuity at the eigenvalues of $H$
and possibly at $E=0$. We can redefine $\xi(E)$ by requiring that $\xi(E+0)=\xi(E)$ for
all $E\in\R$ such that $\xi(E)$ becomes right semicontinuous.

We recall that the spectral shift function is monotone with respect to the perturbation,
i.e. if $H_1$ and $H_2$ are self-adjoint operators such that $H_1\leq H_2$ in the sense of
quadratic forms, then $\xi(E;H_1,H_0)\leq\xi(E;H_2,H_0)$ (see e.g. \cite{Sobolev}).

The value of $\xi(0)$ depends on the spectral properties of the point $E=0$. We call the
point $E=0$ regular iff $(I+V^{1/2}R_0(z)|V|^{1/2})^{-1}$ exists and is bounded at $z=0$.
In the opposite case $E=0$ is called an exceptional point. Here we used the notation
$|V|^{1/2}(x)=|V(x)|^{1/2}$ and $V^{1/2}(x)=\sign V(x)|V(x)|^{1/2}$ and
$R_0(z)=(H_0-z)^{-1}$ is the
resolvent of the free Hamiltonian $H_0$. Other characterizations of the exceptional case
can be found in \cite{BGW,BGK,Aktosun:Klaus:Mee}. The Levinson theorem for Hamiltonians on
a line \cite{Newton,BGW,BGK} states that
\begin{equation}\label{four}
\xi(0)=\xi(+0)=-n_0(V)+\frac{1}{2}
\end{equation}
if $E=0$ is a regular point for $H$ and
\begin{equation}\label{funf}
\xi(0)=\xi(+0)=-n_0(V)
\end{equation}
if $E=0$ is an exceptional point. An extension of (\ref{funf}) for potentials with slower
decrease can be found in \cite{Newton1}.

The scattering matrix $S(E)$ for the pair of Hamiltonians ($H$, $H_0$) at fixed energy
$E\geq 0$ is a 2 by 2 unitary matrix (see \cite{Faddeev,Deift:Trubowitz})
\begin{equation}\label{Smatrix.def}
S(E)=\left(\begin{array}{cc}
           T(E) & R(E) \\
           L(E) & T(E)
           \end{array} \right).
\end{equation}
Below we will use the fact that due to unitarity the S-matrix can be parameterized by the
absolute value of the transmission amplitude $0\leq |T(E)|\leq 1$ and two real valued
phases $\delta(E)$ and $\theta(E)$:
\begin{equation}\label{s.param}
S(E)=\left(\begin{array}{lr}
           |T(E)|e^{i\delta(E)} &  i\sqrt{1-|T(E)|^2}e^{i\delta(E)+i\theta(E)}\\
           i\sqrt{1-|T(E)|^2}e^{i\delta(E)-i\theta(E)}  &
           |T(E)|e^{i\delta(E)}
           \end{array}\right).
\end{equation}
Here $\delta(E)$ is the scattering phase such that $\delta(E)=-\pi\xi(E)$. For reflection
symmetric potentials $\exp\{2i\theta(E)\}\equiv 1$.

Below we will need the following auxiliary results.

\vspace{0.25in}

\bf Lemma 2.1 \it Let $V(x)$ satisfy the condition (\ref{condition}).
Then there is a constant $C_V>0$ such that
\begin{equation}\label{2.1.1}
|\xi(E)|\leq C_V
\end{equation}
for all $E\in\R$\ . Moreover, there is a constant $C>0$ independent of $V$ and $E$ such
that
\begin{equation}\label{2.1.2}
|\xi(E)|\leq C\left\{\frac{1}{2\sqrt{E}}\int_\R|V(x)|dx+\frac{1}{4E}\left[\int_\R |V(x)|
dx\right]^2\right\}
\end{equation}
for all $E>0$. \rm

\vspace{0.25in}

\it Proof. \rm First we prove the inequality (\ref{2.1.2}).
By monotonicity of the spectral shift function we have
\begin{displaymath}
\xi(E;H_0-|V|,H_0)\leq \xi(E)\leq\xi(E;H_0+|V|,H_0).
\end{displaymath}
The operator $|V|^{1/2}R_0(z)|V|^{1/2}$ is trace class for all $z\in\C$ off the positive
real semiaxis (see \cite[Problem 161]{RS4}). This operator has limiting values
$|V|^{1/2}R_0(E\pm i0)|V|^{1/2}$ for all $E>0$ in the Hilbert-Schmidt norm. Since the
resolvent $R_0(z)$ of $H_0$ is an integral operator with kernel
$R_0(x,y;z)=\frac{i}{2\sqrt{z}}e^{i\sqrt{z}|x-y|}$, it follows that
${\Im}|V|^{1/2}R_0(E+i0)|V|^{1/2}$ has the integral kernel
\begin{eqnarray}\label{Einsatz.1}
\lefteqn{\frac{1}{2\sqrt{E}}|V(x)|^{1/2}\cos(\sqrt{E}(x-y))|V(y)|^{1/2}}\nonumber\\
&=&\frac{1}{2\sqrt{E}}|V(x)|^{1/2}\cos(\sqrt{E}x)\cos(\sqrt{E}y)|V(y)|^{1/2}\nonumber\\
&+&\frac{1}{2\sqrt{E}}|V(x)|^{1/2}\sin(\sqrt{E}x)\sin(\sqrt{E}y)|V(y)|^{1/2},
\end{eqnarray}
and thus has a rank 2 and is obviously positive semidefinite.

For positive energies the spectral shift function can be calculated as follows
\cite{Newton,Guillope},
\begin{equation}\label{toestim}
\xi(E;H_0\pm |V|,H_0)=\frac{1}{2\pi i}\log\frac{\det\left(I\pm |V|^{1/2}R_0(E+i0)|V|^{1/2}\right)}
{\det\left(I\pm |V|^{1/2}R_0(E-i0)|V|^{1/2}\right)},
\end{equation}
where the branch of the logarithm is fixed by the condition $\xi(E)\rightarrow 0$ as
$E\rightarrow\infty$. Obviously, the operator $|V|^{1/2}R_0(E\pm i0)|V|^{1/2}$ has no real
eigenvalues (otherwise for some value of the coupling constant $\alpha\in\R$ the number
$-1$ would be an eigenvalue of $\alpha |V|^{1/2}R_0(E\pm i0)|V|^{1/2}$, which implies that
$E\in\sigma_s^+(H_0+\alpha |V|)=\emptyset$, the positive singular part of the spectrum of
$H_0+\alpha |V|$). Therefore, estimating the r.h.s. of (\ref{toestim}) as in
\cite{Sobolev} (Lemmas 2.2, 2.3, and proof of Theorem 3.1) for all $E>0$ we obtain
\begin{eqnarray*}
|\xi(E;H_0\pm|V|,H_0)|\leq C\Big[\|{\Im}|V|^{1/2}R_0(E+i0)|V|^{1/2}\|_{\mathcal{J}_1}\\+
\||V|^{1/2}R_0(E+i0)|V|^{1/2}\|_{\mathcal{J}_2}^2
\Big],
\end{eqnarray*}
where the constant $C$ is independent of $V$ and $E$. Here $\mathcal{J}_1$ and
$\mathcal{J}_2$ denote the trace and Hilbert-Schmidt norms respectively.

Since ${\Im} |V|^{1/2}R_0(E+i0)|V|^{1/2}\geq 0$ is positive semidefinite it follows that
\begin{eqnarray*}
\|{\Im}|V|^{1/2}R_0(E+i0)|V|^{1/2}\|_{\mathcal{J}_1}={\tr}{\Im}|V|^{1/2}R_0(E+i0)|V|^{1/2}\\
=\frac{1}{2\sqrt{E}}\int|V(x)|dx.
\end{eqnarray*}
Obviously, the r.h.s. is also a bound for
$\||V|^{1/2}R_0(E+i0)|V|^{1/2}\|_{\mathcal{J}_2}$.

Now we estimate $\xi(0)$. According to (\ref{four}) and (\ref{funf})
\begin{displaymath}
|\xi(0)|\leq n_0(H)+\frac{1}{2},
\end{displaymath}
where $n_0(H)$ is a total number of eigenvalues of $H$. By the well-known Bargman-type
estimate (see e.g. \cite{RS4})
\begin{displaymath}
n_0(H)\leq 1+\int_\R |x||V(x)|dx
\end{displaymath}
we obtain
\begin{equation}\label{3/2}
|\xi(E)|\leq\frac{3}{2}+\int_\R|x||V(x)|dx
\end{equation}
for $E=0$. Since $\xi(E)$ is a nonincreasing function of $E<0$ and $\xi(E)=0$ for
$E<\inf\sigma(H)$, the estimate (\ref{3/2}) is valid for all $E\leq 0$.

Now let us fix some $E_0>0$. For all $E\notin[0,E_0]$ the estimate
\begin{equation}\label{cv1}
|\xi(E)|\leq C_V^{(1)}
\end{equation}
with
\begin{displaymath}
C_V^{(1)}=\max\left\{\frac{3}{2}+\int_\R |x||V(x)|dx, \frac{C}{2\sqrt{E_0}}\int_\R
|V(x)|dx+\frac{C}{4E_0}\left[\int_\R |V(x)|dx\right]^2\right\}
\end{displaymath}
is valid. The function $\xi(E)$ is continuous on $[0,E_0]$ and thus by the Weierstrass
theorem attains its maximum and its minimum, which by (\ref{2.1.2}) and (\ref{3/2}) are
finite. Therefore, there exists $C_V^{(2)}$ such that $|\xi(E)|\leq C_V^{(2)}$ for all
$E\in[0,E_0]$. This inequality combined with (\ref{cv1}) gives the estimate (\ref{2.1.1})
and completes the proof of the lemma. $\blacksquare$

\vspace{0.25in}

\textit{Remarks:} 1. The first term in the estimate (\ref{2.1.2}) with $C=1/2$ represents
the high-energy asymptotics of the spectral shift function (see e.g.
\cite{Faddeev,Deift:Trubowitz}). Below (Lemma 6.1) we prove that for bounded potentials
$V$ with compact support the second term in (\ref{2.1.2}) can be omitted (with $C$
independent of $V$).

2. The estimate (\ref{2.1.1}) for $E\geq 0$ also follows from the results in
\cite{Schrader}.

\vspace{0.25in}

\bf Lemma 2.2 \it Let $V$ be piecewise continuously differentiable,
satisfy (\ref{condition}) and
\begin{displaymath}
\int_\R (1+|x|)\left|\frac{dV(x)}{dx}\right| dx<\infty.
\end{displaymath}
Then for each closed interval $\Delta\subset\R$
\begin{equation}\label{Jensen}
-\int_\Delta Ed\xi(E) = \tr(\widetilde{V}^{1/2}\EE(\Delta)|\widetilde{V}|^{1/2}),
\end{equation}
where
\begin{displaymath}
\widetilde{V}(x)=V(x)+\frac{1}{2}x\frac{dV(x)}{dx}
\end{displaymath}
and $\EE(\cdot)$ is the spectral resolution for the operator $H$. \rm

\vspace{0.25in}

\textit{Remark:} A similar formula for the case of
operators acting in $L^2(\R^d)$  with $d\geq 2$ and $\Delta\subset(0,+\infty)$ was proved
by Robert and Tamura \cite{Robert:Tamura} and by Jensen \cite{Jensen}.

\vspace{0.25in}

\it Proof. \rm Our proof closely follows the ideas of
\cite{Robert:Tamura}. Let $f_\pm(x,E)$ be the solutions of the integral equations
\begin{eqnarray}
f_+(x,E)=e^{i\sqrt{E}x}-\int_x^{\infty}\frac{\sin\sqrt{E}(x-y)}{\sqrt{E}}V(y)f_+(y,E)dy,
\label{Volt.1}\\
f_-(x,E)=e^{-i\sqrt{E}x}+\int_{-\infty}^x\frac{\sin\sqrt{E}(x-y)}{\sqrt{E}}V(y)f_-(y,E)dy.
\label{Volt.2}
\end{eqnarray}
These functions are solutions of the Schr\"odinger equation
\begin{displaymath}
\left(-\frac{d^2}{dx^2}+V(x)-E\right)f_\pm(x,E)=0.
\end{displaymath}
Let us also consider the functions
\begin{displaymath}
\psi_\pm(x,E)=T(E)f_\pm(x,E),
\end{displaymath}
where $T(E)$ is the transmission amplitude. They satisfy the Lippmann-Schwinger equations
which are integral equations of the Fredholm type \cite{Newton},
\begin{equation}\label{Fredholm}
\psi_\pm(x,E)=\psi_\pm^{(0)}-\frac{i}{2\sqrt{E}}\int_\R e^{i\sqrt{E}|x-y|}V(y)\psi_\pm(y,E)dy
\end{equation}
with
\begin{displaymath}
\psi_\pm^{(0)}(x,E)=e^{\pm i \sqrt{E}x}.
\end{displaymath}
The functions $\psi_\pm(x,E)$ are continuously differentiable with respect to $x\in\R$ and
$E\geq 0$ \cite{Faddeev,Deift:Trubowitz}. Thus,
\begin{displaymath}
\widetilde{\psi}_\pm(x,E;\sigma)=\psi_\pm(x/\sigma,E\sigma^2)
\end{displaymath}
are continuously differentiable with respect to $\sigma$. From (\ref{Fredholm}) and since
\begin{displaymath}
\frac{\partial}{\partial\sigma}\left(\sigma^{-2}V(x/\sigma)\right)_{\sigma=1}=
-2\widetilde{V}(x),
\end{displaymath}
it follows that $\frac{\partial\widetilde{\psi}_\pm}{\partial\sigma}|_{\sigma=1}$ satisfy
the integral equation
\begin{eqnarray*}
\frac{\partial}{\partial\sigma}\widetilde{\psi}_\pm(x,E;\sigma)\Big|_{\sigma=1}=
-\frac{i}{\sqrt{E}}\int_\R e^{i\sqrt{E}|x-y|}\widetilde{V}(y)\psi_\pm(y,E)dy\\
-\frac{i}{2\sqrt{E}}\int_\R e^{i\sqrt{E}|x-y|}V(y)\frac{\partial}{\partial\sigma}
\widetilde{\psi}_\pm(x,E;\sigma)\Big|_{\sigma=1}dy.
\end{eqnarray*}
This easily gives
\begin{equation}\label{dsigma}
\frac{\partial}{\partial\sigma}\widetilde{\psi}_\pm(x,E;\sigma)\Big|_{\sigma=1}=
2\int_\R R(E\pm i0)(x,y)\widetilde{V}(y)\psi_\pm(y,E)dy,
\end{equation}
where $R(z)$ is the resolvent of $-d^2/dx^2+V$.

It is known \cite{Newton} that the scattering matrix (\ref{Smatrix.def}) can be calculated
as
\begin{equation}\label{s.matrix}
S(E)=I-\frac{i}{2\sqrt{E}}\int_\R V(x)
\Psi^{(0)}(x,E)^{\ast}\ \Psi(x,E)dx,
\end{equation}
where
\begin{eqnarray*}
\Psi^{(0)}(x,E) &=& \left(\psi_+^{(0)}(x,E),\psi_-^{(0)}(x,E)\right),\\
\Psi(x,E) &=& \Big(\psi_+(x,E),\psi_-(x,E) \Big),
\end{eqnarray*}
$I$ is the 2 by 2 unit matrix and $\ast$ denotes Hermitian conjugation, such that
\begin{displaymath}
{\Psi^{(0)}}^\ast\Psi=\left(\begin{array}{cc}
           \overline{\psi_+^{(0)}}\psi_+ & \overline{\psi_+^{(0)}}\psi_- \\
           \overline{\psi_-^{(0)}}\psi_+ & \overline{\psi_-^{(0)}}\psi_-\end{array}\right).
\end{displaymath}
Now we calculate
\begin{eqnarray*}
\frac{dS(E)}{dE}&=&\frac{1}{2E}\frac{\partial}{\partial\sigma}S(E\sigma^2)|_{\sigma=1},\\
\frac{\partial}{\partial\sigma}S(E\sigma^2)|_{\sigma=1}&=&-\frac{i}{2\sqrt{E}}
\int_\R\left(\frac{\partial}{\partial\sigma}\sigma^{-2}V(x/\sigma)\right)_{\sigma=1}
\Psi^{(0)}(x,E)^\ast \Psi(x,E)dx\\
&&-\frac{i}{2\sqrt{E}}\int_\R V(x) \Psi^{(0)}(x,E)^\ast \frac{\partial\widetilde{\Psi}}
{\partial\sigma}(x,E;\sigma)|_{\sigma=1} dx.
\end{eqnarray*}
By (\ref{dsigma}) and by the identity
\begin{displaymath}
\psi_\pm(x,E)=\psi_\pm^{(0)}(x,E)-\int_\R R(E\pm i0)(x,y)V(y)\psi_\pm^{(0)}(y,E)dy
\end{displaymath}
we obtain
\begin{displaymath}
\frac{\partial}{\partial\sigma}S(E\sigma^2)\Big|_{\sigma=1}=-\frac{i}{\sqrt{E}}
\int_\R \widetilde{V}(x) \Psi(x,E)^\ast\ \Psi(x,E) dx.
\end{displaymath}
Using the transformation property of the scattering matrix \cite{Newton}
\begin{displaymath}
\left(\begin{array}{c}
      \psi_+(x,E)\\
      \psi_-(x,E)
      \end{array}\right)= S(E)\left(\begin{array}{c}
                                    \overline{\psi_-(x,E)}\\
                                    \overline{\psi_+(x,E)}
                                    \end{array}\right),
\end{displaymath}
we get
\begin{displaymath}
\tr\left(S^\ast(E)\frac{dS(E)}{dE} \right)=
-\frac{i}{2E^{3/2}}\sum_\pm\int_\R\widetilde{V}(x)|\psi_\pm(x,E)|^2 dx.
\end{displaymath}
By the Birman-Krein theorem
\begin{displaymath}
\xi'(E)=\frac{1}{2\pi i}\tr\left(S^\ast(E)\frac{dS(E)}{dE}\right),
\end{displaymath}
and therefore
\begin{displaymath}
\xi'(E)=-\frac{1}{4\pi E^{3/2}}\sum_\pm\int_\R\widetilde{V}(x)|\psi_\pm(x,E)|^2 dx
\end{displaymath}
for all $E>0$.

Now we use the spectral representation for the spectral decomposition of the operator
$-d^2/dx^2+V$ (see e.g. \cite{Gesztesy:Nowell:Potz}),
\begin{eqnarray*}
\EE(x,y,\Delta)=\sum_{j:E_j\in\Delta_-}\psi_j(x)\psi_j(y)\\
+\frac{1}{4\pi}\sum_\pm\int_{\Delta_+}
\psi_\pm(x,E)\overline{\psi_\pm(y,E)}\frac{dE}{\sqrt{E}},
\end{eqnarray*}
where $\Delta_+=\Delta\cap (0,+\infty)$, $\Delta_-=\Delta\cap (-\infty,0)$ and $\psi_j(x)$
are the eigenvectors of $H$ with eigenvalues $E_j<0$.

Since $\widetilde{V}^{1/2}\EE(\Delta)|\widetilde{V}|^{1/2}$ is trace class
\cite{Simon:Semigroup}, we have
\begin{eqnarray*}
\tr\left(\widetilde{V}^{1/2}\EE(\Delta)|\widetilde{V}|^{1/2} \right)=
\sum_{j:E_j\in\Delta_-}\int_\R \widetilde{V}(x)|\psi_j(x)|^2dx-\int_{\Delta_+}E\xi'(E)dE.
\end{eqnarray*}
It remains to prove that
\begin{displaymath}
\sum_{j: E_j\in\Delta_-}\int_\R \widetilde{V}(x)|\psi_j(x)|^2 dx =
-\int_{\Delta_-}Ed\xi(E),
\end{displaymath}
or equivalently
\begin{equation}\label{trace}
\tr\left(\widetilde{V}^{1/2}\EE(\Delta_-)|\widetilde{V}|^{1/2} \right)=
-\int_{\Delta_-}E d\xi(E)=\sum_{j: E_j\in\Delta_-}E_j.
\end{equation}
The proof of (\ref{trace}) is quite elementary. It suffices to consider the case when
$\Delta_-$ contains a single eigenvalue $E_0$ of $H$ with eigenfunction $\phi_0$ such that
$\EE(\Delta_-)$ is the projector onto $\phi_0$.
Let $D$ be the generator of dilations,
\begin{displaymath}
D=(2i)^{-1}\left(x\frac{d}{dx}+\frac{d}{dx}x \right).
\end{displaymath}
In \cite{Jensen:ms} it is shown that
\begin{displaymath}
\widetilde{V}=-\frac{i}{2} [H,D]+H
\end{displaymath}
as a bounded operator from $W^{2,2}(\R)$ to $W^{2,-2}(\R)$, such that
$(\phi_0,\widetilde{V}\phi_0)$ is well defined. Then
\begin{eqnarray*}
\tr(\widetilde{V}^{1/2}\EE(\Delta_-)|\widetilde{V}|^{1/2})=(\phi_0,\widetilde{V}\phi_0)=
E_0-\frac{i}{2}(\phi_0,[H,D]\phi_0)=E_0.
\end{eqnarray*}
$\blacksquare$

\vspace{0.25in}

\textbf{Lemma 2.3} \it For every $z$ with ${\Im} z>0$ the function $\log T(z)$ is analytic and given by
\begin{equation}\label{L.2.3.1}
\log T(z)=-\int_\R\frac{\xi(E)dE}{E-z}.
\end{equation}
Moreover, for all real $\lambda>0$ we have
\begin{equation}\label{L.2.3.2}
\log|T(\lambda)|=\int_\R\log|E-\lambda|d\xi(E),
\end{equation}
where the integral is understood in the Riemann-Stieltjes sense. \rm

\vspace{0.25in}

\textit{Proof.} The analyticity of $\log T(z)$ in the open upper complex half-plane $\C_+$
is well-known (see e.g. \cite{Faddeev}). Also $\log T(z)$ is bounded in $\C_+$ and
\begin{displaymath}
\lim_{\epsilon\rightarrow +0}\log T(E+i\epsilon)=\log|T(E)|-i\pi\xi(E)
\end{displaymath}
in all points of continuity of $\xi(E)$. Therefore, we can apply the Schwarz integral
formula for the half-plane (see e.g. \cite{Schabat}) to reconstruct $\log T(z)$ from the
limiting values of its imaginary part,
\begin{equation}\label{Schwarz}
\log T(z)=-\int_\R\frac{\xi(E)dE}{E-z}+iC,
\end{equation}
with $C$ being a real constant. For $z\rightarrow\infty$ we have
$T(z)=1+\frac{1}{2i\sqrt{z}}\int V(x)dx+O(|z|^{-1})$ \cite{Faddeev,Deift:Trubowitz}.
Therefore, $C$ in (\ref{Schwarz}) must be zero. $\blacksquare$

\vspace{0.25in}

Below we will make use of the Aktosun factorizarion formula \cite{Aktosun}, which we
formulate in the following form.

Let $V(x)$ be some real-valued  locally integrable bounded function. Let
$\{y_n\}_{n\in\Z}$ be a sequence of real numbers such that $y_n\rightarrow\pm\infty$ as
$n\rightarrow\pm\infty$ and $y_n< y_{n+1}$ for all $n\in\Z$. Let $\chi_n(x)$ be the
characteristic function of the interval $[y_n,y_{n+1}]$. We denote $V_n(x)=V(x)\chi_n(x)$
such that
\begin{displaymath}
V^{(-n,m)}(x)=\sum_{j=-n}^m V_j(x)
\end{displaymath}
tends to $V(x)$ as $m,n\rightarrow\infty$. Let $H^{(-n,m)}$ and $H_j$ denote the
Hamiltonians with domains of definition being the Sobolev space $W^{(2,2)}(\R)$,
\begin{displaymath}
H^{(-n,m)}=H_0+V^{(-n,m)},\ H_j=H_0+V_j.
\end{displaymath}
Let $S^{(-n,m)}(E)$ and $S_j(E)$ be the corresponding S-matrices,
\begin{displaymath}
S^{(-n,m)}(E)=\left(\begin{array}{cc}
                    T^{(-n,m)}(E) & R^{(-n,m)}(E) \\
                    L^{(-n,m)}(E) & T^{(-n,m)}(E)
                    \end{array}\right),
\end{displaymath}
\begin{displaymath}
S_j(E)=\left(\begin{array}{cc}
                    T_j(E) & R_j(E) \\
                    L_j(E) & T_j(E)
                    \end{array}\right).
\end{displaymath}
We also consider the matrices
\begin{displaymath}
\Lambda^{(-n,m)}(E)=\left(\begin{array}{lr}
        \frac{1}{T^{(-n,m)}(E)}  & -\frac{R^{(-n,m)}(E)}{T^{(-n,m)}(E)} \\
        \frac{L^{(-n,m)}(E)}{T^{(-n,m)}(E)}  & \frac{1}{T^{(-n,m)}(E)^\ast}
        \end{array}\right)
\end{displaymath}
and
\begin{displaymath}
\Lambda_j(E)=\left(\begin{array}{lr}
        \frac{1}{T_j(E)}  & -\frac{R_j(E)}{T_j(E)} \\
        \frac{L_j(E)}{T_j(E)}  & \frac{1}{T_j(E)^\ast}
        \end{array}\right).
\end{displaymath}
From the unitarity of the scattering matrix (see e.g. \cite{Faddeev}) it follows that the
matrices $\Lambda^{(-n,m)}(E)$ and $\Lambda_j(E)$ are unimodular. We note that in
Faddeev's terminology \cite{Faddeev} the elements of $\Lambda(E)$ are given  by the
coefficients $c_{ij}(\sqrt{E})$. More precisely,
\begin{displaymath}
\begin{array}{ll}
\displaystyle \frac{1}{T(E)}=c_{12}(\sqrt{E}), &
\displaystyle -\frac{R(E)}{T(E)}=-c_{22}(\sqrt{E}), \\
\displaystyle \frac{L(E)}{T(E)}=c_{11}(\sqrt{E}), &
\displaystyle \frac{1}{T(E)^\ast}=c_{12}(-\sqrt{E}).
\end{array}
\end{displaymath}
The Aktosun factorization formula states that
\begin{equation}\label{Aktosun.form}
\Lambda^{(-n,m)}(E)=\prod_{j=-n}^m \Lambda_j(E).
\end{equation}
Here and below we understand the product $\prod$ in the ordered sense, i.e.
\begin{displaymath}
\prod_{j=-n}^m \Lambda_j(E)=\Lambda_{-n}(E)\cdots\Lambda_m(E).
\end{displaymath}

The theorem below provides an alternative proof of (\ref{Aktosun.form}). Actually we show
that the factorization property of the matrices $\Lambda$ is directly related to the
propagator property of the fundamental solution of the corresponding Schr\"odinger
equation.

\vspace{0.25in}

\textbf{Theorem 2.4} \it For arbitrary $E>0$ consider the matrix $U(x,x';E)\in\SL(2;\mathbb{C})$
which solves the initial value problem
\begin{equation}\label{U}
\frac{dU(x,x';E)}{dx}=-\frac{i}{2\sqrt{E}}V(x)M(x,E)U(x,x';E),\ U(x,x;E)=I,
\end{equation}
with
\begin{displaymath}
M(x,E)=\left(\begin{array}{cc}
             1 & e^{-2i\sqrt{E}x} \\
             -e^{2i\sqrt{E}x} & -1
             \end{array}\right).
\end{displaymath}
The matrices $\Lambda^{(-n,m)}(E)$ are related to $U(x,x';E)$ such that
\begin{equation}\label{theor.2.3}
\Lambda^{(-n,m)}(E)=U(y_{-n},y_m;E).
\end{equation}
\rm

\vspace{0.25in}

The factorization formula (\ref{Aktosun.form}) follows immediately from (\ref{theor.2.3})
and from the propagator property of $U$, $U(x,x'';E)U(x'',x';E)=U(x,x';E)$.

\vspace{0.25in}

\textit{Remarks}: 1. The matrix $U(x,x';E)$ is related to the fundamental solution
$\phi(x,x';E)$ of the Schr\"odinger equation
\begin{equation}\label{Schroed}
-\frac{d^2u}{dx^2}+V(x)u-Eu=0,
\end{equation}
which satisfies
\begin{displaymath}
\frac{d}{dx}\phi(x,x';E)=\left(\begin{array}{cc}
                               0 & 1\\
                               V(x)-E & 0\end{array}\right)
                               \phi(x,x';E),\ \phi(x,x;E)=I.
\end{displaymath}
It is easy to see that
\begin{displaymath}
\phi(x,x';E)=P(x,E)U(x,x';E)P(x',E)^{-1},
\end{displaymath}
where
\begin{displaymath}
P(x,E)=\left(\begin{array}{cc}
             e^{i\sqrt{E}x} & e^{-i\sqrt{E}x} \\
             i\sqrt{E}e^{i\sqrt{E}x} & -i\sqrt{E}e^{-i\sqrt{E}x}
             \end{array}\right),
\end{displaymath}
such that
\begin{displaymath}
\left(\begin{array}{cc}
                               0 & 1\\
                               V(x)-E & 0\end{array}\right)=
\frac{dP(x,E)}{dx}P(x,E)^{-1}+\frac{1}{2i\sqrt{E}}V(x)P(x,E)M(x,E)P(x,E)^{-1}.
\end{displaymath}

2. The formula (\ref{theor.2.3}) reduces the problem of the study of $\phi(y_{-n},y_m;E)$
to the study of the scattering matrix for the corresponding single-site potential. Note
that $P(x,E)^\ast P(x,E)$ is not a multiple of the identity operator. Therefore, the
Hilbert-Schmidt norms of $\Lambda^{(-n,m)}(E)$ and of $\phi(y_{-n},y_m;E)$ are not equal.
This will become relevant in Section 5.

3. The connection between the solutions of (\ref{U}) and scattering characteristics was
noted earlier in \cite{Marchenko:Pastur1,Papanicolaou}.

\vspace{0.25in}

\textit{Proof.} To prove the theorem it suffices to consider $V(x)$ supported on the interval
$[0,a]$ and to show that $U(x,E):=U(x,x';E)|_{x'=0}$, i.e. the solution of the equation
\begin{displaymath}
\frac{dU(x,E)}{dx}=-\frac{i}{2\sqrt{E}}V(x)M(x,E)U(x,E),\ U(0,E)=I,
\end{displaymath}
satisfies
\begin{displaymath}
U(a,E)=\Lambda(E)=\left(\begin{array}{cc}
                        \frac{1}{T(E)} & -\frac{R(E)}{T(E)} \\
                        \frac{L(E)}{T(E)} & \frac{1}{T(E)^\ast}
                        \end{array}\right),
\end{displaymath}
where $T(E)$, $R(E)$, and $L(E)$ correspond to the potential $V$.

Consider the solutions $\psi_\pm(x,E)$ of the equation (\ref{Schroed}), such that
\begin{displaymath}
\psi_+(x,E)=\left\{\begin{array}{ll}
                   e^{i\sqrt{E}x}+R(E)e^{-i\sqrt{E}x}, & x<0, \\
                   T(E)e^{i\sqrt{E}x}, & x>a,
                   \end{array}\right.
\end{displaymath}
\begin{displaymath}
\psi_-(x,E)=\left\{\begin{array}{ll}
                   T(E)e^{-i\sqrt{E}x}, & x<0,\\
                   e^{-i\sqrt{E}x}+L(E)e^{i\sqrt{E}x}, & x>a.
                   \end{array}\right.
\end{displaymath}
We introduce the 2 by 2 matrix
\begin{displaymath}
W(x,E)=\left(\begin{array}{cc}
             A_-(x,E) & A_+(x,E) \\
             B_-(x,E) & B_+(x,E)
             \end{array}\right),
\end{displaymath}
where $A_\pm(x,E)$ and $B_\pm(x,E)$ are given by
\begin{eqnarray*}
A_\pm(x,E)=\frac{1}{2}e^{i\sqrt{E}x}\left[\psi_\pm(x,E)+\frac{i}{\sqrt{E}}
\frac{d\psi_\pm(x,E)}{dx} \right],\\
B_\pm(x,E)=\frac{1}{2}e^{-i\sqrt{E}x}\left[\psi_\pm(x,E)-\frac{i}{\sqrt{E}}
\frac{d\psi_\pm(x,E)}{dx} \right].
\end{eqnarray*}
It is easy to see that $W(x,E)$ satisfies the equation
\begin{equation}\label{W}
\frac{dW(x,E)}{dx}=-\frac{i}{2\sqrt{E}}V(x)M(x,E)W(x,E),
\end{equation}
and
\begin{displaymath}
W(0,E)=\left(\begin{array}{cc}
              T(E) & R(E) \\
              0 & 1
              \end{array}\right).
\end{displaymath}
Also we have
\begin{displaymath}
W(a,E)=\left(\begin{array}{cc}
               1 & 0 \\
              L(E) & T(E)
              \end{array}\right).
\end{displaymath}
Obviously, $W(x,E)W(0,E)^{-1}$ also satisfies (\ref{W}) and equals $I$ for $x=0$. Thus,
\begin{displaymath}
U(x,E)=W(x,E)W(0,E)^{-1}.
\end{displaymath}
Moreover,
\begin{displaymath}
U(a,E)=\left(\begin{array}{cc}
               1 & 0 \\
              L(E) & T(E)\end{array}\right)\left(\begin{array}{cc}
              T(E) & R(E) \\
              0 & 1 \end{array}\right)^{-1}=\Lambda(E),
\end{displaymath}
since
\begin{displaymath}
T(E)-\frac{R(E)L(E)}{T(E)}=\frac{1}{T(E)^\ast}.
\end{displaymath}
Finally note that $U(x,x';E)$, defined by (\ref{U}), is given as
$U(x,x';E)=U(x,E)U(x',E)^{-1}$. $\blacksquare$

\section{Cluster Property of the Spectral Shift Function}
\setcounter{equation}{0}
\markright{3 Cluster Property of the Spectral Shift Function}

In this Section we establish a cluster property of the spectral shift function for
Schr\"{o}dinger operators in $L^2(\R)$ of the form
\begin{equation}\label{2.1}
H(d)=-\frac{d^2}{dx^2}+V_d,\ V_d=V_1+V_2(\cdot-d),
\end{equation}
where the $V_i$ are in $L^1$ and have compact supports (see also
\cite{KoSch1,KoSch2,KoSch3} for results concerning cluster properties in the higher
dimensional case). We study the behavior of the spectral shift function $\xi(E;d)$ for the
pair ($H(d)$, $H_0$) when $|d|$ is sufficiently large. More precisely let
$\mathcal{D}=\mathcal{D}(V_1,V_2)\subset\R$ be such that the intersection of the minimal
closed intervals containing $\supp V_1$ and $\supp V_2(\cdot-d)$ is at most a point. This
implies that $V_1(x)V_2(x-d)=0$ a.e. for all $d\in\mathcal{D}$. We will henceforth assume
that $d\in\mathcal{D}$. Denote
\begin{equation}\label{3.3.neu}
\xi_{12}(E;d)=\xi(E;H(d),H_0)-\xi(E;H_1,H_0)-\xi(E;H_2,H_0)
\end{equation}
with $H_i=H_0+V_i$  ($i=1,2$). Also we set $H_2(d)=H_0+V_2(\cdot-d)$ such that
$H_2=H_2(d=0)$. By the translation invariance of the spectral shift function, we have
$\xi(E; H_2(d), H_0)=\xi(E;H_2,H_0)$ for all $d$. For brevity in what follows we will
write $\xi(E;d)=\xi(E;H(d),H_0)$, $\xi_i(E)=\xi(E;H_i,H_0)$, $i=1,2$.

Below we will need the following simple result:

\vspace{0.25in}

\bf Lemma 3.1 \it Suppose that $H^{(1)}$, $H_0^{(1)}$ and  $H^{(2)}$, $H_0^{(2)}$ are
semi-bounded self-adjoint operators in the Hilbert spaces $\mathcal{H}^{(1)}$ and
$\mathcal{H}^{(2)}$ respectively, such that $H^{(i)}-H_0^{(i)}$, $i=1,2$ are trace class.
Then for a.e. $E\in\R$
\begin{eqnarray}\label{0.1}
\lefteqn{\xi(E;H^{(1)}\oplus H^{(2)},H_0^{(1)}\oplus H_0^{(2)})=}\nonumber\\
&&\xi(E;H^{(1)}\oplus H_0^{(2)},H_0^{(1)}\oplus H_0^{(2)})+
\xi(E;H_0^{(1)}\oplus H^{(2)},H_0^{(1)}\oplus H_0^{(2)}).
\end{eqnarray}
Moreover,
\begin{eqnarray}\label{0.2}
\xi(E;H^{(1)}\oplus H_0^{(2)},H_0^{(1)}\oplus H_0^{(2)})=\xi(E;H^{(1)},H_0^{(1)}),\nonumber\\
\xi(E;H_0^{(1)}\oplus H^{(2)},H_0^{(1)}\oplus H_0^{(2)})=\xi(E;H^{(2)},H_0^{(2)})
\end{eqnarray}
a.e. on $\R$.\rm

\vspace{0.25in}

\it Proof. \rm Let $\mathcal{H}=\mathcal{H}^{(1)}\oplus \mathcal{H}^{(2)}$. For every
$f\in C_0^\infty(\R)$ we have
\begin{eqnarray*}
\lefteqn{{\tr}_\mathcal{H}\left(f(H^{(1)}\oplus H^{(2)})-f(H_0^{(1)}\oplus H_0^{(2)}) \right)}\\
&=& {\tr}_\mathcal{H}\left(f(H^{(1)})\oplus f(H^{(2)})-f(H_0^{(1)})\oplus
f(H_0^{(2)})\right)\\ &=& {\tr}_{\mathcal{H}^{(1)}}\left(f(H^{(1)})-f(H_0^{(1)}) \right) +
{\tr}_{\mathcal{H}^{(2)}}\left(f(H^{(2)})-f(H_0^{(2)}) \right)\\ &=&{\tr}_\mathcal{H}
\left(f(H^{(1)})\oplus f(H_0^{(2)})- f(H_0^{(1)})\oplus f(H_0^{(2)})\right)
\\&&+
{\tr}_\mathcal{H}
\left(f(H_0^{(1)})\oplus f(H^{(2)})- f(H_0^{(1)})\oplus f(H_0^{(2)})\right)\\
&=& {\tr}_\mathcal{H}\left(f(H^{(1)}\oplus H_0^{(2)})-f(H_0^{(1)}\oplus H_0^{(2)})
\right)\\
&&+{\tr}_\mathcal{H}\left(f(H_0^{(1)}\oplus H^{(2)})-f(H_0^{(1)}\oplus H_0^{(2)})
\right).
\end{eqnarray*}
From this it follows that (\ref{0.1}) and (\ref{0.2}) are valid up to an additive
constant. From our normalization and from the semiboundedness of the operators involved it
follows that this constant equals zero. $\blacksquare$

\vspace{0.25in}

\bf Theorem 3.2 \it For all $d\in\mathcal{D}(V_1,V_2)$
\begin{displaymath}
|\xi_{12}(E;d)|\leq\left\{\begin{array}{ll}
                          3/2, & E\geq 0 \\
                          1,   & E<0.
                          \end{array}\right.
\end{displaymath}

\vspace{0.25in}

\it Proof. \rm Let us fix some $y\in\R$ lying between the supports of $V_1$ and $V_2(\cdot-d)$.
More precisely we require that $y$ is such that $\supp V_1\subset(-\infty,y]$ and $\supp
V_2(\cdot-d)\subset[y,\infty)$ if $d>0$ and $\supp V_2(\cdot-d)\subset(-\infty,y]$ and
$\supp V_1\subset[y,\infty)$ if $d<0$. Without loss of generality we may suppose that
$d>0$. For every potential $V$ satisfying (\ref{condition}) along with the Hamiltonian $H$
we consider the self-adjoint Schr\"odinger operators on $L^2(\R)$
\begin{displaymath}
H^{(D,N)}=-\frac{d^2}{dx^2}+V
\end{displaymath}
with Dirichlet and Neumann boundary conditions at $x=y$. In accordance with the
decomposition
\begin{displaymath}
L^2(\R)=L^2(-\infty,y)\oplus L^2(y,\infty)
\end{displaymath}
one can write $H^{(D,N)}=H_-^{(D,N)}\oplus H_+^{(D,N)}$. From Krein's formula (see e.g.
\cite{Combes:Duclos:Seiler}) it follows that $H^{(D)}$ and $H^{(N)}$ are rank one
perturbations of $H$. Also we have
\begin{equation}\label{ineq.ND}
H^{(N)}\leq H\leq H^{(D)}
\end{equation}
in the sense of quadratic forms. Since $V_1$ and $V_2(\cdot-d)$ have disjoint compact
supports, we have
\begin{displaymath}
H^{(D,N)}(d)=H_{1,-}^{(D,N)}\oplus H_{2,+}^{(D,N)}(d).
\end{displaymath}

From Lemma 3.1 it follows that
\begin{eqnarray}\label{xi.prove}
\lefteqn{\xi(E;H^{(D,N)},H_0^{(D,N)})=
\xi(E;H_{1,-}^{(D,N)}\oplus H_{2,+}^{(D,N)}(d),H_0^{(D,N)})}\nonumber\\
& = & \xi(E;H_1^{(D,N)},H_0^{(D,N)})+\xi(E;H_2^{(D,N)}(d),H_0^{(D,N)}).
\end{eqnarray}

Now by the chain rule for the spectral shift function \cite{BiYa} and from
(\ref{xi.prove}) it follows that
\begin{eqnarray}\label{xi.1}
\lefteqn{\xi(E;H(d),H_0)=\xi(E;H(d),H^{(D,N)}(d))}\nonumber\\ &&
+\xi(E;H^{(D,N)}(d),H_0^{(D,N)})+\xi(E;H_0^{(D,N)},H_0)\nonumber\\ &=&
\xi(E;H(d),H^{(D,N)}(d))+\xi(E;H_0^{(D,N)},H_0)\nonumber\\ && +
\xi(E;H_1^{(D,N)},H_0^{(D,N)})+\xi(E;H_2^{(D,N)}(d),H_0^{(D,N)}).
\end{eqnarray}
On the other hand and again by the chain rule one has
\begin{eqnarray}\label{xi.2}
\lefteqn{\xi(E;H_i,H_0)=\xi(E;H_1,H_1^{(D,N)})}\nonumber\\
&&+\xi(E;H_1^{(D,N)},H_0^{(D,N)})+\xi(E;H_0^{(D,N)},H_0)
\end{eqnarray}
and analogously with $H_1$ replaced by $H_2(d)$. From (\ref{xi.1}) and (\ref{xi.2}) it
follows that
\begin{eqnarray*}
\xi(E;H(d),H_0)=\xi(E;H_1,H_0)+\xi(E;H_2(d),H_0)\\
+\xi(E;H(d),H^{(D,N)}(d))-\xi(E;H_0^{(D,N)},H_0)\\-
\xi(E;H_1,H_1^{(D,N)})-\xi(E;H_2(d),H_2^{(D,N)}(d))
\end{eqnarray*}
such that
\begin{eqnarray}\label{ineq.long}
\lefteqn{\xi_{12}(E;d)=-\xi(E;H^{(D,N)}(d),H(d))-\xi(E;H_0^{(D,N)},H_0)}\nonumber\\
&& +\xi(E;H_1^{(D,N)},H_1)+\xi(E;H_2^{(D,N)}(d),H_2(d)).
\end{eqnarray}
We note that in the terminology of Gesztesy and Simon \cite{Gesztesy:Simon}
$\xi(E;H^{(D)}(d),H(d))$,\newline $\xi(E;H_i^{(D)},H_i)$, $i=1,2$, and
$\xi(E;H_0^{(D)},H_0)$ are the xi-functions for the operators $H(d)$, $H_i$, $i=1,2$, and
$H_0$ respectively.

From (\ref{ineq.ND}) and from the fact that the absolute value of the spectral shift
function for rank one perturbations is not greater than one, we have that for all real $E$
\begin{eqnarray*}
0\leq\xi(E;H^{(D)},H)\leq 1,\\
-1\leq\xi(E;H^{(N)},H)\leq 0,
\end{eqnarray*}
where $H$ stands for one of the operators $H(d)$, $H_1$, $H_2(d)$, or $H_0$. Also (see
\cite{Gesztesy:Simon})
\begin{eqnarray*}
\xi(E;H_0^{(D)},H_0)&=&\left\{\begin{array}{ll}
                            1/2, & E\geq 0\\
                            0,   & E<0
                            \end{array}\right. ,\\
\xi(E;H_0^{(N)},H_0)&=&\left\{\begin{array}{ll}
                            -1/2, & E\geq 0\\
                            0,   & E<0
                            \end{array}\right. .
\end{eqnarray*}
Therefore, from (\ref{ineq.long}) we obtain
\begin{eqnarray*}
-3/2\leq \xi_{12}(E;d)\leq 3/2, && E\geq 0,\\
-1\leq \xi_{12}(E;d)\leq 2, && E< 0
\end{eqnarray*}
using Dirichlet boundary conditions and
\begin{eqnarray*}
-3/2\leq \xi_{12}(E;d)\leq 3/2, && E\geq 0,\\
-2\leq \xi_{12}(E;d)\leq 1, && E< 0
\end{eqnarray*}
using Neumann boundary conditions. This completes the proof of the theorem. $\blacksquare$

\vspace{0.25in}

For our further purposes to be elaborated below Theorem 3.2 provides a completely
sufficient information. However, we note that when $E\geq 0$ the bound for $\xi_{12}(E;d)$
can be improved:

\vspace{0.25in}

\bf Theorem 3.3 \it For all $d\in\mathcal{D}(V_1,V_2)$ and all $E\geq 0$
\begin{equation}\label{3.4.neu}
|\xi_{12}(E;d)|\leq 1/2.
\end{equation}
\rm

\vspace{0.25in}

The proof of Theorem 3.3 will given below.

\vspace{0.25in}

Theorems 3.2 and 3.3 imply that $|\xi_{12}(E;d)|\leq 1$ for all $d\in\mathcal{D}(V_1,V_2)$
and all $E\in\R$. Let us define the functions
\begin{eqnarray}\label{3.6}
\widetilde{\xi}^{(\pm)}(E,d) &=& \xi(E;d)\pm 1,\nonumber\\
  \\
\widetilde{\xi}_j^{(\pm)}(E) &=& \xi_j(E)\pm 1,\ j=1,2.\nonumber
\end{eqnarray}

\vspace{0.25in}

\bf Corollary 3.4 \it For all $E\in\R$ the functions (\ref{3.6}) satisfy the
inequalities
\begin{eqnarray}
\widetilde{\xi}^{(+)}(E;d) \leq \widetilde{\xi}_1^{(+)}(E)+
\widetilde{\xi}_2^{(+)}(E),\label{3.2.1}\\
\widetilde{\xi}^{(-)}(E;d) \geq \widetilde{\xi}_1^{(-)}(E)+
\widetilde{\xi}_2^{(-)}(E),\label{3.2.2}
\end{eqnarray}
i.e. $\widetilde{\xi}^{(+)}(E)$ is a subadditive and
$\widetilde{\xi}^{(-)}(E)$ is a superadditive function with respect
to the potentials $V_1$ and $V_2(\cdot-d)$.  \rm

\vspace{0.25in}

\it Proof. \rm Due to (\ref{3.3.neu}) and Theorems 3.2, 3.3  we have
\begin{displaymath}
\xi(E;d)=\xi_1(E)+\xi_2(E)+\xi_{12}(E;d)\leq\xi_1(E)+\xi_2(E)+1.
\end{displaymath}
Adding $1$ to both sides of this inequality we arrive at
(\ref{3.2.1}). Similarly, we have
\begin{displaymath}
\xi(E;d)=\xi_1(E)+\xi_2(E)+\xi_{12}(E;d)\geq\xi_1(E)+\xi_2(E)-1.
\end{displaymath}
Subtracting $1$ from both sides yields (\ref{3.2.2}). $\blacksquare$

\vspace{0.25in}

\it Proof of Theorem 3.3. \rm Recall (see \cite{Faddeev,Deift:Trubowitz}) that the
scattering matrix at energy $E\geq 0$ for the pair of Hamiltonians
($H_j=H_0+V_j$, $H_0$) is given by
\begin{displaymath}
S_j(E)=\left(\begin{array}{lr}
             T_j(E) & R_j(E) \\
             L_j(E) & T_j(E)
             \end{array}\right),\ j=1,2.
\end{displaymath}
We use the fact that for all $d\in\mathcal{D}(V_1,V_2)$ and all $E>0$
\begin{equation}\label{3.5}
\xi_{12}(E;d)=-\frac{1}{2\pi i}
\log\frac{1-R_1(E)L_2(E)e^{2i\sqrt{E}d}}
{1-R_1(E)^\ast L_2(E)^\ast e^{-2i\sqrt{E}d}}.
\end{equation}
The formula (\ref{3.5}) has appeared earlier in \cite{Bianchi} and follows from the
Aktosun factorization formula \ref{Aktosun.form}
\begin{equation}\label{Aktosun.1}
T(E;d)=\frac{T_1(E)T_2(E)}{1-R_1(E)L_2(E)e^{2i\sqrt{E}d}}
\end{equation}
and the identity (see e.g. \cite{Newton})
\begin{equation}\label{phase}
\frac{T_j(E)}{T_j(E)^\ast}=\det S_j(E) =e^{-2\pi i\xi_j(E)},\ j=1,2.
\end{equation}

According to (\ref{Smatrix.def}) the reflection amplitudes $L_j(E)$ and $R_j(E)$ can be
represented in the form
\begin{displaymath}
L_j(E)=A_j(E)e^{i\delta_j^{(L)}},\quad R_j(E)=A_j(E)e^{i\delta_j^{(R)}}
\end{displaymath}
with $0\leq A_j(E)\leq 1$. Moreover $A_j(E)=1$ only when $T_j(E)=0$. This in turn can
happen only if $E=0$ (see \cite{Faddeev,Deift:Trubowitz}). Therefore
\begin{eqnarray*}
&&\log\frac{1-R_1(E)L_2(E)e^{2i\sqrt{E}d}}{1-R_1(E)^\ast L_2(E)^\ast
e^{-2i\sqrt{E}d}}\\
&&=\log\frac{1-A_1(E)A_2(E)e^{i(\delta_1^{(R)}+\delta_2^{(L)}+2\sqrt{E}d)}}
{1-A_1(E)A_2(E)e^{-i(\delta_1^{(R)}+\delta_2^{(L)}+2\sqrt{E}d)}}\\
&&=-2i\arctan\frac{A_1(E)A_2(E)\sin(\delta_1^{(R)}+\delta_2^{(L)}+
2\sqrt{E}d)}{1-A_1(E)A_2(E)\cos(\delta_1^{(R)}+\delta_2^{(L)}+
2\sqrt{E}d)}.
\end{eqnarray*}
Due to the fact that $0\leq A_j(E)< 1$ for $E>0$ the absolute value of this expression is
strictly bounded by $\pi$. $\blacksquare$

\vspace{0.25in}

Since the spectral shift function is right continuous the estimate
\begin{equation}\label{Eeq0}
|\xi_{12}(E;d)|\leq 1/2
\end{equation}
holds also for $E=0$. We note that the inequality (\ref{Eeq0}) for $E=0$ also follows from
the Levinson theorem \cite{BGW,BGK}, Theorem 3.1 of \cite{Klaus} and the results of
\cite{Aktosun:Klaus:Mee}. We include this discussion (see also \cite{Bianchi}) to show the
different aspects of the problem. For brevity we set $n(V)=n_0(V)$, the number of bound
states below $E=0$. Theorem 3.1 of \cite{Klaus} states in particular that
$n(V_1+V_2(\cdot-d))=n(V_1)+n(V_2(\cdot-d))=n(V_1)+n(V_2)$ for all $d$ such that
$d\in\mathcal{D}(V_1,V_2)$ if $E=0$ is an exceptional point for at least one of $H_1$ or
$H_2$. If $E=0$ is a regular point for both $H_1$ and $H_2$ then
\begin{displaymath}
n(V_1)+n(V_2)-1\leq n(V_1+V_2(\cdot-d)) \leq n(V_1)+n(V_2).
\end{displaymath}
Therefore, by (\ref{four}) and (\ref{funf}) we have
\begin{displaymath}
-\xi_1(0)-\xi_2(0)\leq n(V_1+V_2(\cdot-d))\leq -\xi_1(0)-\xi_2(0)+1
\end{displaymath}
if $E=0$ is a regular point for both $H_1$ and $H_2$, and
\begin{displaymath}
n(V_1+V_2(\cdot-d))=-\xi_1(0)-\xi_2(0)+1/2
\end{displaymath}
if $E=0$ is an exceptional point for at least one of the operators $H_1$ and $H_2$. Then
according to Theorem 2.3 of \cite{Aktosun:Klaus:Mee} we have the alternatives

(i) Let $E=0$ be an exceptional point for exactly one of the operators $H_1$ and $H_2$.
Then $E=0$ is a regular point for $H(d)$ and therefore
\begin{displaymath}
\xi(0;d)=\xi_1(0)+\xi_2(0).
\end{displaymath}

(ii) Let $E=0$ be an exceptional point for both operators $H_1$ and $H_2$. Then $E=0$ is
an exceptional point for $H(d)$ and therefore
\begin{displaymath}
\xi(0;d)=\xi_1(0)+\xi_2(0)-1/2.
\end{displaymath}

Let now $E=0$ be a regular point for both $H_1$ and $H_2$. The discussion at the end of
Section 2 in \cite{Aktosun:Klaus:Mee} states that $E=0$ is a regular point for $H(d)$ for
all except possibly one value $d=d_0\in\R$\ . Excluding this value $d_0$ (however, we
found no proof that $d_0\notin\mathcal{D}(V_1,V_2)$) we get that
\begin{displaymath}
\xi_1(0)+\xi_2(0)-\frac{1}{2}\leq \xi(0;d) \leq \xi_1(0)+\xi_2(0)+\frac{1}{2}.
\end{displaymath}
for all $d\in\mathcal{D}(V_1,V_2)$ with $d\neq d_0$.

\vspace{0.25in}

Now we reconsider the case $E<0$. We recall that in this case the spectral shift
function equals minus the number of eigenvalues less than $E$. The discussion below shows
that the bound $|\xi_{12}(E;d)|\leq 1$ for $E<0$ in general cannot be improved.

Let $E_{k_i}^{(i)}<0$, $k_i=1,\ldots, n_0(V_i)$, $i=1,2$ be the eigenvalues of $H_i$. Let
$E_k(d)<0$, $k=1,\ldots,n_0(V_1+V_2(\cdot-d))$ be the eigenvalues of $H(d)$. We recall
some known properties of $E_k(d)$ \cite{Klaus}. The functions $E_k(d)$ are real analytic
in $d\in\mathcal{D}=\mathcal{D}(V_1,V_2)$. First let $E<0$ be an eigenvalue of both $H_1$
and $H_2$. Then $H(d)$ has two eigenvalues $E_\pm(d)$, $E_-(d)<E<E_+(d)$, which both
converge to $E$ as $d\rightarrow\infty$. Moreover, $E_-^{\prime}>0$ and $E_+^{\prime}<0$
(with prime denoting the derivative with respect to $d$) for all $d\in\mathcal{D}$.
Secondly, let $E$ be an eigenvalue of (say) $H_1$ but not of $H_2$. Then the only
eigenvalue $E(d)$ of $H(d)$ approaches $E$ as $d\rightarrow\infty$. Moreover, either
$E'(d)>0$, or $E'(d)<0$, or $E'(d)=0$ for all $d\in\mathcal{D}$. There are no eigenvalues
of $H(d)$ other than the ones described above. We note also that for all $d\in\mathcal{D}$
either $E_k(d)\neq E_k^{(i)}$ for any $k, k_i$ and $i$, or $E_k(d)=E_{k_i}^{(i)}$ for some
$k, k_i$ and $i$. Indeed let us suppose that there is $d_0\in\mathcal{D}$ such that
$E_k(d_0)=E_{k_i}^{(i)}=E_0$ for some $k, k_i$ and $i$. Then inspecting the proof of
Theorem 1.2 in \cite{Klaus} we see that this implies $E_k(d)=E_0$ for all
$d\in\mathcal{D}$. As is well known in the one-dimensional case the eigenvalues of $H(d)$
are simple and therefore the $E_k(d)$'s cannot cross each other.

Now fix some $E_0<0$. There are two cases to be considered: (i) $E_0$ is an eigenvalue
of neither $H_1$ nor $H_2$, (ii) $E_0$ is an eigenvalue of at least one of the
Hamiltonians $H_1$, $H_2$. By the discussion above in case (i) there is $d(E_0)>0$
such that for all $d\geq d(E_0)$ one has
$n_{E_0}(V_1+V_2(\cdot-d))=n_{E_0}(V_1)+n_{E_0}(V_2)$. When
$d\in\mathcal{D}$ decreases only one of the curves $E_k(d)$ can pass through $E_0$ (since
the $E_k(d)$'s cannot cross each other) thus decreasing or increasing
$n_{E_0}(V_1+V_2(\cdot-d))$ by one. In case (ii) $n_{E_0}(V_1+V_2(\cdot-d))$ does not
depend on $d\in\mathcal{D}$ and equals $n_{E_0}(V_1)+n_{E_0}(V_2)$ or differs from
$n_{E_0}(V_1)+n_{E_0}(V_2)$ by one.

\section{Existence of  the Spectral Shift Density}
\setcounter{equation}{0}
\markright{4 Existence of  the Spectral Shift Density}

As stated in the Introduction we will consider random Schr\"{o}dinger operators
$H(\omega)$ in $L^2(\R)$ of the form (\ref{1.1}) with $\{\alpha_j(\omega)\}_{j\in{\Z}}$
being a sequence of i.i.d. variables on a probability space $(\Omega,{\cal F},\P)$ having
a common density $\varphi$, which is continuous and has support in the finite interval
$[\alpha_-,\alpha_+]\subset\R$. Also the sequence $\{\alpha_j(\omega)\}_{j\in{\Z}}$  is
supposed to form a stationary, metrically transitive random field, i.e. there are measure
preserving, ergodic transformations $\{T_j\}_{j\in{\Z}}$ on $\Omega$ such that
$\alpha_j(T_k\omega)=\alpha_{j-k}(\omega)$.

We suppose that the single-site potential $f$ is in $C(\R)$ with $\supp f\subseteq
[-1/2,1/2]$ and $f\geq 0$.

First we introduce the Hamiltonians
\begin{equation}\label{Hamiltonian}
H^{(-n,m)}(\omega)=-\frac{d^2}{dx^2}+\sum_{j=-n}^{m}\alpha_j(\omega) f(\cdot-j),
\end{equation}
such that $H^{(n)}(\omega)=H^{(-n,n)}(\omega)$. Let $\xi^{(-n,m)}(E;\omega)$ be the
spectral shift function for the pair $(H^{(-n,m)}(\omega),H_0)$.

The operators $H^{(-n,m)}(T_k\omega)$ and  $H^{(-n-k,m-k)}(\omega)$
are unitarily equivalent. In fact, consider the unitary shift operator
$U_j,\ j\in\Z$ on $L^2(\R)$ given as $U_jf(x)=f(x-j)$. Then one has
\begin{eqnarray}\label{transl}
H^{(-n,m)}(T_k\omega)=-\frac{d^2}{dx^2}+\sum_{j=-n}^m \alpha_j(T_k\omega)
f(\cdot-j)\nonumber\\
=-\frac{d^2}{dx^2}+\sum_{j=-n}^m \alpha_{j-k}(\omega)f(\cdot-j)
\nonumber\\
=-\frac{d^2}{dx^2}+\sum_{j=-n-k}^{m-k} \alpha_{j}(\omega)f(\cdot-j-k)
\nonumber\\
=U_k H^{(-n-k,m-k)}(\omega)U_k^{\ast}.
\end{eqnarray}
Since the spectral shift functions for pairs of unitarily
equivalent operators are equal, we have
\begin{displaymath}
\xi^{(-n,m)}(E;T_k\omega)=\xi^{(-n-k,m-k)}(E;\omega).
\end{displaymath}
This remains true for the functions
\begin{displaymath}
\xi^{(-n,m)}_\pm(E;\omega)=\xi^{(-n-k,m-k)}(E;\omega)\pm 1.
\end{displaymath}

Now let $k$ be an arbitrary integer such that $-n\leq k<m$. Then due to Corollary 3.4 we
have that
\begin{equation}\label{subadditiv}
\xi^{(-n,m)}_+(E;\omega) \leq \xi^{(-n,k)}_+(E;\omega)+
\xi^{(k,m)}_+(E;\omega)
\end{equation}
and
\begin{displaymath}
\xi^{(-n,m)}_-(E;\omega) \geq \xi^{(-n,k)}_-(E;\omega)+
\xi^{(k,m)}_-(E;\omega).
\end{displaymath}
Now we show that
\begin{eqnarray*}
\Gamma_+&=&\inf_{m,n}\frac{1}{m+n+1}\E\left\{\xi^{(-n,m)}_+(E;\omega) \right\}
> -\infty,\\
\Gamma_-&=&\sup_{m,n}\frac{1}{m+n+1}\E\left\{\xi^{(-n,m)}_-(E;\omega) \right\}
<\infty,
\end{eqnarray*}
where $\E$ denotes the expectation with respect to the probability measure
$\P$.

First we note that
\begin{eqnarray*}
\Gamma_- &=& \sup_{m,n}\frac{1}{(m+n+1)}\E\left\{\xi_-^{(-n,m)}(E;\omega) \right\}\\
&=& \sup_{m,n}\frac{1}{(m+n+1)}\E\left\{\xi_+^{(-n,m)}(E;\omega)-2 \right\}\\ &\leq&
\sup_{m,n}\frac{1}{(m+n+1)}\E\left\{\xi_+^{(-n,m)}(E;\omega) \right\}.
\end{eqnarray*}
From the inequality (\ref{subadditiv}) it follows that
\begin{displaymath}
\xi_+^{(-n,m)}(E;\omega)\leq \sum_{j=-n}^m \xi(E;H_0+\alpha_j(\omega)f,H_0)+(n+m+1)
\end{displaymath}
and then by the monotonicity theorem of the spectral shift function with respect to
perturbations we have
\begin{displaymath}
\xi_+^{(-n,m)}(E;\omega)\leq (n+m+1)\left[\xi(E;H_0+\alpha_+ f,H_0)+1 \right].
\end{displaymath}
Hence by Lemma 2.1
\begin{displaymath}
\Gamma_-\leq \xi(E;H_0+\alpha_+ f,H_0)+1<\infty.
\end{displaymath}

Similarly we can prove that $\Gamma_+> -\infty$. Indeed,
\begin{eqnarray*}
\Gamma_+ &=& \inf_{m,n}\frac{1}{m+n+1}\E\left\{\xi_+^{(-n,m)}(E;\omega) \right\}\\
&=& \inf_{m,n}\frac{1}{m+n+1}\E\left\{\xi_-^{(-n,m)}(E;\omega)+2 \right\}\\
&\geq&\inf_{m,n}\frac{1}{m+n+1}\E\left\{\xi_-^{(-n,m)}(E;\omega) \right\}\\
&\geq&\inf_{m,n}\frac{1}{m+n+1}\sum_{j=-n}^m\left(\E\left\{\xi(E;H_0+\alpha_j(\omega),H_0)
\right\}-1\right)\\
&\geq& \xi(E;H_0+\alpha_-f,H_0)-1>-\infty.
\end{eqnarray*}
Thus we have proved

\vspace{0.25in}

\bf Theorem 4.1 \it For every $E\in\R$ the family $\xi_+^{(-n,m)}(E;\omega)$ is a
subadditive and $\xi_-^{(-n,m)}(E;\omega)$ is a superadditive random process. \rm

\vspace{0.25in}

Applying now the Akcoglu-Krengel superadditive ergodic theorem we
obtain that for every $E\in\R$ there is a set $\Omega_E\subset\Omega$
of full measure such that
\begin{equation}\label{limit}
\lim_{m,n\rightarrow\infty}\frac{\xi^{(-n,m)}(E;\omega)}{n+m+1}=:\xi(E)
\end{equation}
exists and is non-random. We call this limit the spectral shift density.

Now the problem is to show that the set $\Omega_E$ of full measure
can be chosen to be independent of $E$ as long as $E$ is a point of
continuity of the limit. We note that a priori the set $\cap_{E\in\R}\Omega_E$
is not necessarily of full measure.

We recall how this problem is solved for the density of states
$N(E)$ (see e.g. \cite[p. 312]{CaLa}). Once one has established the
existence of the limit
\begin{equation}\label{den.states}
\lim_{m,n\rightarrow\infty}\frac{N_\omega^{(-n,m)}(E)}{n+m+1}=N(E)
\end{equation}
for every fixed $E$ and $\P$-almost all $\omega\in\Omega$, one can
choose $\widetilde{\Omega}$ as the intersection of all sets
$\Omega_E$ when $E$ runs through the rationals and redefine the
limiting function $N(E)$ to make it right continuous. Since $N(E)$
is a monotone nondecreasing function of $E$, this could change the
values of the limiting function on at most a countable set of
discontinuities. Hence (\ref{den.states}) is valid at every
continuity point of $N(E)$ for all $\omega\in\widetilde{\Omega}$,
which is obviously of full measure.

In our case the limiting function $\xi(E)$ is not monotone. However intuition says that
$\xi(E)$ must be equal to $N_0(E)-N(E)$ (this is indeed the case as will be proven below).
Therefore, $\xi(E)$ is expected to be at least of bounded variation. The simplest way to
prove this is to show that $(n+m+1)^{-1}\xi_\omega^{(-n,m)}(E)$ are Lipshitz functions
with Lipshitz constants bounded uniformly in $n$ and $m$, which is nothing but a
Wegner-type estimate \cite{Wegner} for the spectral shift density. This guarantees that
$\xi(E)$ is Lipshitz continuous.

We will need the spectral averaging theorem \cite{BiSo,Simon:preprint}:

\vspace{0.25in}

\bf Lemma 4.2 \it Let $H=H_0+W$ with $W\in L^1(\R)$. Let $V\in L^1(\R)$ be nonnegative.
$\EE_s(\cdot)$ the spectral decomposition of unity for $H_s=H+sV$. Then for any Borel set
$\Delta\subset\R$
\begin{displaymath}
\int_{s_0}^{s_1}\tr(V^{1/2}\EE_s(\Delta)V^{1/2})ds=\int_\Delta\xi(E;H_{s_1},H_{s_0})dE.
\end{displaymath}
\rm

\vspace{0.25in}

\textit{Remark:} $\tr (V^{1/2}E_s(\Delta)V^{1/2})$ is well defined since for every $f\in L^1(\R)$
the operator $f^{1/2}E_s(\Delta)|f|^{1/2}$ is trace class (see \cite{Simon:Semigroup}).

\vspace{0.25in}

The present formulation of the spectral averaging theorem is a direct consequence of a
slightly extended version of Theorem 4 in \cite{Simon:preprint}. This extension is
straightforward and therefore we do not discuss details here. Now we prove

\vspace{0.25in}

\bf Theorem 4.3 \it Let $f$ be piecewise continuously differentiable
with $\supp f\subseteq[-1/2,1/2]$ and assume there is a constant $c_f>0$ such that
$$\left|f(x)+\frac{x}{2}\frac{df}{dx}\right|\leq c_f f(x)$$ for a.e. $x\in\supp f$. Then
$E\xi(E)$ is Lipshitz continuous for all $E\in\R$, i.e. for each closed interval
$\Delta\subset\R$ there is a constant $C_\Delta$ such that
\begin{displaymath}
|E_2\xi(E_2)-E_1\xi(E_1)|\leq C_\Delta |E_2-E_1|
\end{displaymath}
for all $E_1, E_2\in\Delta$. \rm

\vspace{0.25in}

Thus it suffices to determine $\xi(E)$ for $E$ running over a
Lebesgue-dense set of $\R$, say the rationals. We choose the set
$\widetilde{\Omega}$ as the intersection $\cap_{E\in\Q}\Omega_E$.
The limit (\ref{limit}) then exists for all $E\in\R$ and all
$\omega\in\widetilde{\Omega}$.

Now we turn to the proof of Theorem 4.3. We note that the method of the proof is not
restricted to the one-dimensional case and can be extended to higher dimensions (see
\cite{KoSch3}).

\vspace{0.25in}

\it Proof of Theorem 4.3. \rm We apply Lemma 2.2 to $H=H^{(-n,m)}(\omega)$.
Let us set $\Delta_1=[E_1,E_2]\subseteq\Delta$. Then we have
\begin{eqnarray*}
\lefteqn{E_2\xi^{(-n,m)}(E_2;\omega)-E_1\xi^{(-n,m)}(E_1;\omega)}\\
&=&\int_{\Delta_1}\xi^{(-n,m)}(E;\omega)dE+\int_{\Delta_1}Ed\xi^{(-n,m)}(E;\omega)\\
&=&\int_{\Delta_1}\xi^{(-n,m)}(E;\omega)dE
-\sum_{j=-n}^m\alpha_j(\omega)\tr\left(\widetilde{f}^{1/2}(\cdot-j)
\EE_\omega^{(-n,m)}(\Delta_1)|\widetilde{f}|^{1/2}(\cdot-j) \right)\\
&=&\int_{\Delta_1}\xi^{(-n,m)}(E;\omega)dE
-\sum_{j=-n}^m\alpha_0(T_{-j}\omega)\tr\left(\widetilde{f}^{1/2}
\EE_{T_{-j}\omega}^{(-n-j,m-j)}(\Delta_1)|\widetilde{f}|^{1/2} \right),
\end{eqnarray*}
where
\begin{displaymath}
\widetilde{f}=f+\frac{1}{2}x\frac{df}{dx},
\end{displaymath}
and $\EE_\omega^{(-n,m)}(\cdot)$ is the spectral resolution for
$H^{(-n,m)}(\omega)$. Therefore
\begin{eqnarray}\label{3.terms}
&&\E\left\{E_2\xi^{(-n,m)}(E_2;\omega)-E_1\xi^{(-n,m)}(E_1;\omega) \right\}
\nonumber\\
&=&\E\left\{\int_{\Delta_1}\xi^{(-n,m)}(E;\omega) dE
\right\}
-\sum_{j=-n}^m\E\left\{\alpha_0(\omega)\tr
\left(\widetilde{f}^{1/2}\EE_\omega^{(-n-j,m-j)}(\Delta_1)
|\widetilde{f}|^{1/2} \right) \right\}.\qquad\quad
\end{eqnarray}
First let us estimate the second term on the r.h.s. of (\ref{3.terms}),
\begin{eqnarray*}
\left|\E\left\{\alpha_0(\omega)\tr
\left(\widetilde{f}^{1/2}\EE_\omega^{(-n-j,m-j)}(\Delta_1)
|\widetilde{f}|^{1/2} \right) \right\} \right|\\
\leq \alpha_+\E\left\{\left|\tr\left(\widetilde{f}^{1/2}
\EE_\omega^{(-n-j,m-j)}(\Delta_1)|\widetilde{f}|^{1/2} \right) \right| \right\}.
\end{eqnarray*}
Since for any $A\in\mathcal{J}_1$ the inequality $|\tr A|\leq\tr|A|$ holds, we have
\begin{eqnarray*}
\left|\tr\left(\widetilde{f}^{1/2}
\EE_\omega^{(-n-j,m-j)}(\Delta_1)|\widetilde{f}|^{1/2} \right) \right|  \leq
\tr\left(|\widetilde{f}|^{1/2}\EE_\omega^{(-n-j,m-j)}(\Delta_1) |\widetilde{f}|^{1/2} \right)\\
\leq c_f\tr\left(f^{1/2}\EE_\omega^{(-n-j,m-j)}(\Delta_1) f^{1/2}\right).
\end{eqnarray*}
Let us denote
\begin{displaymath}
H_\alpha^{(-n,m)}(\omega)=H_0+\sum_{\substack{j=-n\\ j\neq 0}}^m\alpha_j(\omega)
f(\cdot-j)+\alpha f,
\end{displaymath}
and let $\EE_{\omega,\alpha}^{(-n,m)}(\cdot)$ be the corresponding
resolution of the identity. Then we have
\begin{eqnarray*}
\lefteqn{\E\left\{\tr\left(f^{1/2}\EE_\omega^{(-n-j,m-j)}(\Delta_1)f^{1/2}\right)\right\}}\\
&=&\E\left\{\int_{\alpha_-}^{\alpha_+}d\alpha\phi(\alpha)
\tr\left(f^{1/2}\EE_{\omega,\alpha}^{(-n-j,m-j)}(\Delta_1) f^{1/2}\right)\right\}\\
&\leq& \|\phi\|_\infty \E\left\{\int_{\alpha_-}^{\alpha_+}d\alpha
\tr\left(f^{1/2}\EE_{\omega,\alpha}^{(-n-j,m-j)}(\Delta_1) f^{1/2}\right)\right\}.
\end{eqnarray*}
Now we apply Lemma 4.2, according to which we obtain
\begin{eqnarray*}
\int_{\alpha_-}^{\alpha_+}d\alpha\ \tr\left(f^{1/2}
\EE_{\omega,\alpha}^{(-n-j,m-j)}(\Delta_1) f^{1/2}\right)\\
=\int_{\Delta_1} dE\ \xi(E;H_{\alpha_+}^{(-n-j,m-j)}(\omega),
H_{\alpha_-}^{(-n-j,m-j)}(\omega)),
\end{eqnarray*}
where $\xi(E;H_{\alpha_+}^{(-n-j,m-j)}(\omega), H_{\alpha_-}^{(-n-j,m-j)}(\omega))$ stands
for the spectral shift function of the pair ($H_{\alpha_+}^{(-n-j,m-j)}(\omega)$,
$H_{\alpha_-}^{(-n-j,m-j)}(\omega)$). By the chain rule and Corollary 3.4 we have
\begin{eqnarray*}
\lefteqn{\xi(E;H_{\alpha_+}^{(-n-j,m-j)}(\omega), H_{\alpha_-}^{(-n-j,m-j)}(\omega))}\\
&=& \xi(E;H_{\alpha_+}^{(-n-j,m-j)}(\omega),H_0)-
\xi(E;H_{\alpha_-}^{(-n-j,m-j)}(\omega),H_0)\\
&\leq&
\xi(E;H_{\alpha=0}^{(-n-j,m-j)}(\omega),H_0)+\xi(E;H_0+\alpha_+f,H_0)+1\\ &&
-\xi(E;H_{\alpha=0}^{(-n-j,m-j)}(\omega),H_0)-\xi(E;H_0+\alpha_-f,H_0)+1\\ &=&
\xi(E;H_0+\alpha_+f,H_0+\alpha_-f)+2.
\end{eqnarray*}
Therefore, the second term on the r.h.s. of (\ref{3.terms}) can be bounded by
\begin{equation}\label{bra.1}
(n+m+1)|E_2-E_1|\|\phi\|_\infty
\left[\max_{E\in\Delta_1}\xi(E;H_0+\alpha_+f,H_0+\alpha_-f)+2 \right].
\end{equation}
By Lemma 2.1 and since
$\xi(E;H_0+\alpha_+f,H_0+\alpha_-f)=\xi(E;H_0+\alpha_+f,H_0)-\xi(E;H_0+\alpha_-f,H_0)$ the
maximum of $\xi(E;H_0+\alpha_+f,H_0+\alpha_-f)$ is bounded.

Now we estimate the first term on the r.h.s. of (\ref{3.terms}). By the Fubini theorem
\begin{eqnarray*}
\lefteqn{\E\left\{\int_{\Delta_1}\xi^{(-n,m)}(E;\omega)dE\right\}}\\
&=&\int_{\Delta_1}\E\left\{\xi^{(-n,m)}(E;\omega) \right\}dE.
\end{eqnarray*}
Using monotonicity and Corollary 3.4 we obtain
\begin{equation}\label{bra.2}
\E\left\{\xi^{(-n,m)}(E_1;\omega)\right\}
\leq (n+m+1)\left[\xi(E_1;H_0+\alpha_+ f,H_0) +1\right].
\end{equation}
The expressions in the square brackets in (\ref{bra.1}) and (\ref{bra.2}) are finite by
Lemma 2.1.

Thus we have proved that
\begin{eqnarray}\label{limit.vor}
(n+m+1)^{-1}\left|E_2\E\left\{\xi^{(-n,m)}(E_2;\omega)
\right\}-E_1\E\left\{\xi^{(-n,m)}(E_1;\omega)\right\}\right|\nonumber\\
\leq C_\Delta|E_2-E_1|.
\end{eqnarray}
Now we note that by the Lebesgue dominated convergence theorem
\begin{displaymath}
\lim_{m,n\rightarrow\infty}(n+m+1)^{-1}
\E\left\{\xi_\omega^{(-n,m)}(E)\right\}=\xi(E)
\end{displaymath}
for every fixed $E\in\R$. Thus, taking the limit $n,m\rightarrow\infty$ in
(\ref{limit.vor}) we arrive at the claim of theorem. $\blacksquare$

\vspace{0.25in}

\bf Theorem 4.4 \it The spectral shift density is the difference of the integrated
densities of states for the free and the interacting Hamiltonians, $\xi(E)=N_0(E)-N(E)$
for all $E\in\R$.
\rm

\vspace{0.25in}

\it Proof. \rm Let $D^{(-n,m)}(\omega)$ and $D_0^{(-n,m)}$ be the self-adjoint Schr\"odinger
operators corresponding to the differential expression (\ref{Hamiltonian}) and
$H_0=-d^2/dx^2$ respectively on $L^2(-n-1/2,m+1/2)$ with Dirichlet boundary conditions at
$x=-n-1/2$ and $x=m+1/2$ (the Friedrichs extension of (\ref{Hamiltonian}) on
$C_0^\infty(-n-1/2,m+1/2)$). They have purely discrete spectrum and therefore the spectral
shift function $\xi(E;D^{(-n,m)}(\omega),D_0^{(-n,m)})$ is simply the difference of the
corresponding counting functions
\begin{displaymath}
N(E;D_0^{(-n,m)})-N(E;D^{(-n,m)}(\omega)).
\end{displaymath}
It is well known (see e.g. \cite{CaLa}) that for all $E\in\R$
\begin{eqnarray*}
N(E)=\lim_{m,n\rightarrow\infty}\frac{N(E;D^{(-n,m)}(\omega))}{n+m+1},\\
N_0(E)=\lim_{m,n\rightarrow\infty}\frac{N(E;D_0^{(-n,m)})}{n+m+1}=\sqrt{E}/\pi
\end{eqnarray*}
almost surely. Hence
\begin{equation}\label{N0NE}
N_0(E)-N(E)=\lim_{m,n\rightarrow\infty}\frac{\xi(E;D^{(-n,m)}(\omega),D_0^{(-n,m)})}{n+m+1}
\end{equation}
for almost all $\omega\in\Omega$.

Now we prove that the difference
\begin{displaymath}
\xi(E;H^{(-n,m)}(\omega),H_0)-\xi(E;D^{(-n,m)}(\omega),D_0^{(-n,m)})
\end{displaymath}
is bounded in absolute value by 2 uniformly in $n,m\in\R$ and $\omega\in\Omega$.
Therefore, the existence of the limit (\ref{N0NE}) immediately implies the existence of
$\xi(E)$ and the equality $\xi(E)=N_0(E)-N(E)$. Thus, in fact we do not need Theorem 4.3
in this context.

By the chain rule for the spectral shift function we have
\begin{eqnarray}\label{discuss}
\xi(E;H^{(-n,m)}(\omega),H_0) = -\xi(E;H_D^{(-n,m)}(\omega),H^{(-n,m)}(\omega))\nonumber\\
+\xi(E;H_D^{(-n,m)}(\omega),H_{0,D}^{(-n,m)})+\xi(E;H_{0,D}^{(-n,m)},H_0).
\end{eqnarray}
Here $H_D^{(-n,m)}(\omega)$ denotes the operator (\ref{Hamiltonian}) on $L^2(\R)$ with
Dirichlet boundary conditions at $x=-n-1/2$ and $x=m+1/2$ such that
\begin{equation}\label{decomposition}
H_D^{(-n,m)}(\omega)=D_0^{(-\infty,-n)}\oplus D^{(-n,m)}(\omega) \oplus D_0^{(m,\infty)}
\end{equation}
with respect to the direct decomposition of the Hilbert space
$L^2(\R)=L^2(-\infty,-n-1/2)\oplus L^2(-n-1/2,m+1/2)\oplus L^2(m+1/2,\infty)$. Similarly
there is a decomposition of
\begin{displaymath}
H_{0,D}^{(-n,m)}=-\frac{d^2}{dx^2}
\end{displaymath}
with the same boundary conditions. From Krein's formula it follows that
$H_D^{(-n,m)}(\omega)$ and $H_{0,D}^{(-n,m)}$ are rank two perturbations of
$H^{(-n,m)}(\omega)$ and $H_0$, respectively. Thus, we immediately have
\begin{eqnarray}\label{improve}
0\leq \xi(E;H_D^{(-n,m)}(\omega),H^{(-n,m)}(\omega))\leq 2,\nonumber\\ 0\leq
\xi(E;H_{0,D}^{(-n,m)},H_0)\leq 2
\end{eqnarray}
for all $n,m\in\N$ and $\omega\in\Omega$. Actually, (\ref{improve}) can be improved
\cite[Remark 5.2]{Jensen:Kato} such that
\begin{displaymath}
\frac{1}{2}\leq \xi(E;H_{0,D}^{(-n,m)},H_0)\leq\frac{3}{2}.
\end{displaymath}

From (\ref{decomposition}) by Lemma 3.1 it follows that
\begin{equation}\label{gleich}
\xi(E;H_D^{(-n,m)}(\omega),H_{0,D}^{(-n,m)})=\xi(E;D^{(-n,m)}(\omega),D_0^{(-n,m)}).
\end{equation}
This completes the proof of the theorem. $\blacksquare$

\vspace{0.25in}

\textit{Remark:} We comment on the formula (\ref{discuss}). We note that
$\xi(E;H^{(-n,m)}(\omega),H_0)$ is
continuous in $E>0$ although $\xi(E;H_D^{(-n,m)}, H_{0,D}^{(-n,m)})$ is a step-like
function being the difference of two counting functions. This means that the difference
\begin{displaymath}
\xi(E;H_{0,D}^{(-n,m)},H_0))-\xi(E;H_D^{(-n,m)}(\omega),H^{(-n,m)}(\omega))
\end{displaymath}
compensates the jumps of $\xi(E;H_D^{(-n,m)}(\omega),H_{0,D}^{(-n,m)})$ making
the r.h.s. of (\ref{discuss}) continuous. This was noted by Jensen and Kato
\cite{Jensen:Kato}. A similar phenomenon was found for obstacle scattering in
$\R^2$ by Eckmann and Pillet \cite{Eckmann:Pillet}.

\centerline{\epsfig{figure=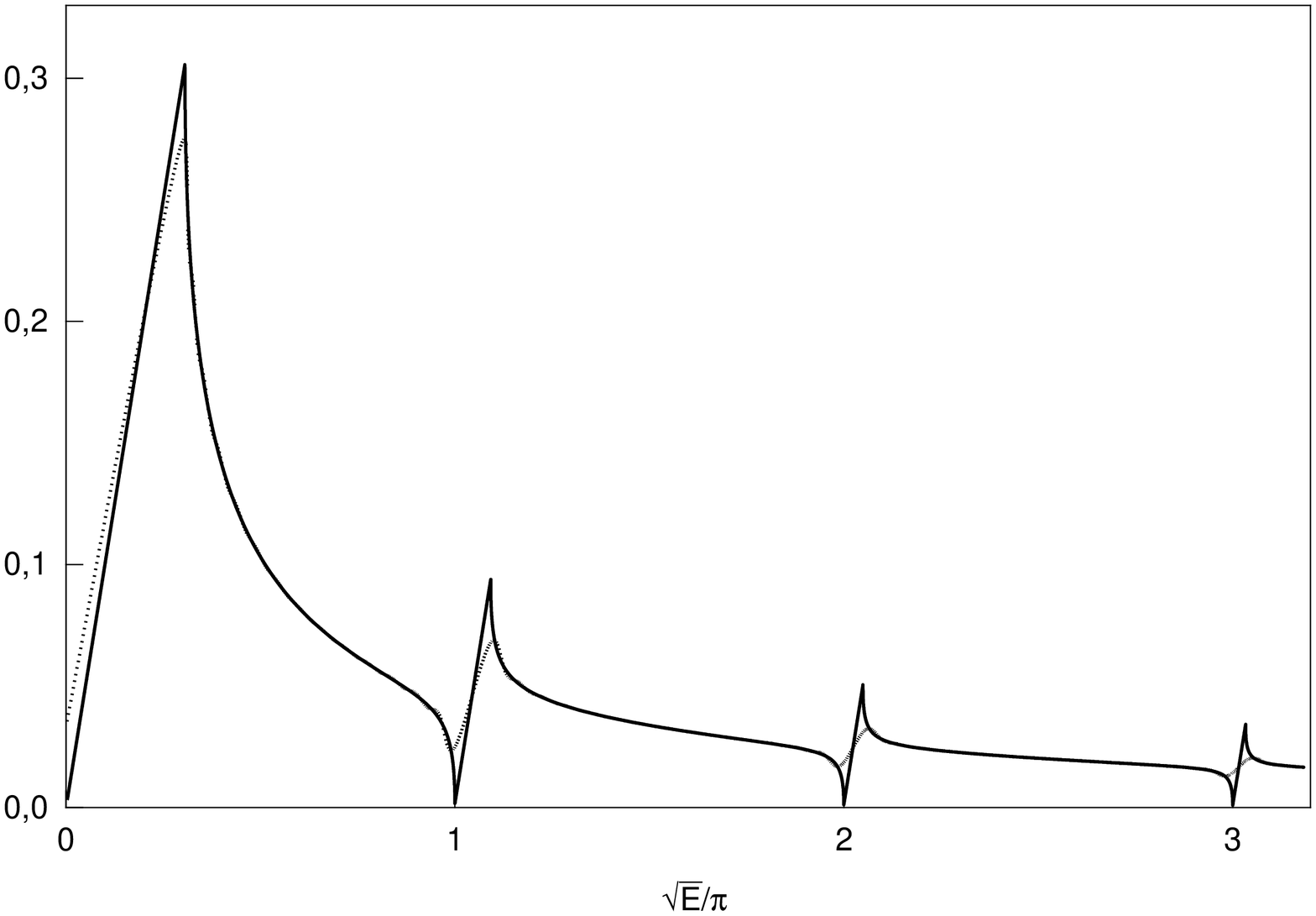,width=18cm,clip=}}
\centerline{Fig. 1: Spectral shift density for the deterministic Kronig-Penney model
(see text).}

\vspace{0.25in}

As an illustration to Theorem 4.4 (see Fig. 1) we have calculated
$(n+m+1)^{-1}\xi^{(-n,m)}(E)$ (dotted line) for $n=m=7$ with $f$ taken to be
the point interaction and $\alpha_j(\omega)\equiv 1$ (the Kronig-Penney model)
and compared this result with $N_0(E)-N(E)$ (solid line). The density of states
$N(E)$ for this case can be given in closed analytic form (see e.g.
\cite{Albeverio:book}).

\section{Density of the Transmission Coefficient\\ and the Lyapunov Exponent}
\setcounter{equation}{0}
\markright{5 Lyapunov Exponent}

Now we turn to a discussion of the density of the transmission
coefficient. Let $T_\omega^{(-n,m)}(E)$ be the transmission
amplitude at energy $E>0$ for the Hamiltonian $H^{(-n,m)}(\omega)$
(\ref{Hamiltonian}). We now prove

\vspace{0.25in}

\bf Theorem 5.1 \it For every fixed $E>0$ and almost all $\omega\in\Omega$
the limit
\begin{displaymath}
\lim_{n,m\rightarrow\infty}\frac{\log |T_\omega^{(-n,m)}(E)|}{n+m+1}
=:-\gamma_T(E)
\end{displaymath}
exists and is non-random.
\rm

\vspace{0.25in}

\textit{Remark:} For periodic deterministic potentials the behavior of $T^{(-n,n)}(E)$
as $n\rightarrow\infty$ was studied numerically in \cite{Rorres}.

\vspace{0.25in}

A very similar statement was proven earlier by Marchenko and Pastur
\cite{Marchenko:Pastur1}. Our proof is a slight modification of
that given in \cite{Marchenko:Pastur1}. We start with

\vspace{0.25in}

\bf Lemma 5.2 \it  For all $E\geq 0$ the transmission amplitude for the Hamiltonian
(\ref{2.1}) satisfies the inequality
\begin{displaymath}
|T(E)|\geq \frac{1}{2}|T_1(E)|\ |T_2(E)|.
\end{displaymath}
\rm

\vspace{0.25in}

\it Proof. \rm By the Aktosun factorization formula
\begin{displaymath}
T(E)=T_1(E) T_2(E)\left(1-R_1(E)L_2(E)\right)^{-1}.
\end{displaymath}
By the unitarity of the scattering matrix $|R_j(E)|\leq 1$  and
$|L_j(E)|\leq 1$, $j=1,2$  for all $E\geq 0$. Hence
\begin{displaymath}
|1-R_1(E)L_2(E)|\leq 2
\end{displaymath}
and the claim follows. $\blacksquare$

\vspace{0.25in}

\it Proof of Theorem 5.1 \rm We set
\begin{displaymath}
t_\omega^{(-n,m)}(E)=\frac{1}{2}\log|T_\omega^{(-n,m)}(E)|\leq 0 .
\end{displaymath}
From the fact that
\begin{displaymath}
T^{(-n,m)}_{T_k\omega}(E)=T^{(-n-k,m-k)}_\omega(E),\ k\in\Z,
\end{displaymath}
which is is an immediate consequence of (\ref{transl}) we obtain
\begin{displaymath}
t^{(-n,m)}_{T_k\omega}(E)=t^{(-n-k,m-k)}_\omega(E).
\end{displaymath}
By Lemma 5.2 we have
\begin{displaymath}
t_\omega^{(-n,k)}(E)+t_\omega^{(k,m)}(E)
\leq t_\omega^{(-n,m)}(E)
\end{displaymath}
for all $k$ with $-n<k<m$. Thus $t_\omega^{(-n,m)}(E)$  is a superadditive random process.
Theorem 5.1 now follows by Akcoglu-Krengel superadditive ergodic theorem. $\blacksquare$

\vspace{0.25in}

\bf Theorem 5.3 \it For every $E>0$
$\gamma_T(E)$ equals $\gamma(E)$, the upper Lyapunov exponent for the fundamental
matrix of the Schr\"{o}dinger operator.
\rm

\vspace{0.25in}

This observation (although without a complete proof) is known (see
\cite{Lifshitz:Gredeskul:Pasur,Lifshitz:book}). We start with recalling the definition of
the Lyapunov exponent (see e.g. \cite{Figotin:Pastur}). Let $\phi_\omega(x;E)$ be the
fundamental matrix of the Schr\"{o}dinger equation
\begin{equation}\label{Schr}
-\frac{d^2\psi}{dx^2}+\sum_{j\in\Z}\alpha_j(\omega)f(\cdot-j)\psi
=E\psi
\end{equation}
such that
\begin{displaymath}
\left(\begin{array}{c}
      u(x) \\ u'(x) \end{array}\right)=\phi_\omega(x;E)
\left(\begin{array}{c}
      u(0) \\ u'(0) \end{array}\right)
\end{displaymath}
for any solution of (\ref{Schr}). The Lyapunov exponents $\gamma_\omega^\pm(E)$ are
defined by
\begin{equation}\label{Lyap.def}
\gamma_\omega^\pm(E)=\underset{x\rightarrow\pm\infty}{\overline{\lim}}
\frac{1}{|x|}\log\|\phi_\omega(x;E)\|.
\end{equation}
Actually it can be shown that $\gamma_\omega^+(E)=\gamma_\omega^-(E)=
\gamma(E)$ for fixed $E$ and $\P$-almost all $\omega$. Moreover,
$\overline{\lim}$ can be replaced by $\lim$ (see e.g. \cite{Figotin:Pastur}).

Now fix $E>0$. We have to show that for $\P$-almost all $\omega$
\begin{displaymath}
\gamma(E)=\lim_{x\rightarrow\pm\infty}\frac{1}{|x|}\log\|\phi_\omega(x;E) \|
=\gamma_T(E).
\end{displaymath}
First let us redefine the fundamental matrix such that
\begin{displaymath}
\left(\begin{array}{c}
      u(x+1/2) \\ u'(x+1/2) \end{array}\right)=\widetilde{\phi}_\omega(x;E)
\left(\begin{array}{c}
      u(1/2) \\ u'(1/2) \end{array}\right).
\end{displaymath}
This obviously does not change the Lyapunov exponent, which we can calculate
as a limit over integer $x$,
\begin{displaymath}
\gamma(E)=\lim_{n\rightarrow\infty}\frac{1}{n}
\log\|\widetilde{\phi}_\omega(n;E)\|.
\end{displaymath}
This implies that we can express $\widetilde{\phi}_\omega(n;E)$ in terms of the
transmission and reflection amplitudes for the Hamiltonian $H^{(1,n)}(\omega)$. More
precisely, let us consider the particular solutions $\psi^{(\pm)}(x;E)$ of the
Schr\"{o}dinger equation with Hamiltonian  $H^{(1,n)}(\omega)$ corresponding to the energy
$E>0$ and for $x\leq 1/2$ having the form
\begin{displaymath}
\psi^{(\pm)}_\omega(x;E)=e^{\pm i\sqrt{E}x}.
\end{displaymath}
Then it is easy to see (see \cite{Faddeev} and Section 2), that for $x\geq n+1/2$ these
solutions have the form
\begin{eqnarray*}
\psi^{(-)}_\omega(x;E) &=& \frac{L_\omega^{(1,n)}(E)}{T_\omega^{(1,n)}(E)}e^{i\sqrt{E}x}
+\frac{1}{T_\omega^{(1,n)}(E)}e^{-i\sqrt{E}x},\\ \\
\psi^{(+)}_\omega(x;E) &=& \overline{\psi^{(-)}_\omega(x;E)},
\end{eqnarray*}
where $T_\omega^{(1,n)}(E)$ and $L_\omega^{(1,n)}(E)$ are transmission and reflection
amplitudes for the Hamiltonian $H^{(1,n)}(\omega)$, respectively. Therefore, the matrix
elements of $\widetilde{\phi}_\omega(n;E)$ are given by
\begin{eqnarray*}
\left[\widetilde{\phi}_\omega(n;E)\right]_{11} &=& \frac{1}{2}
\left[\frac{L_\omega^{(1,n)}(E)^\ast}{T_\omega^{(1,n)}(E)^\ast}
e^{-i\sqrt{E}(n+1)}+
\frac{L_\omega^{(1,n)}(E)}{T_\omega^{(1,n)}(E)}e^{i\sqrt{E}(n+1)}\right]\\
&& +\frac{1}{2}\left[\frac{e^{i\sqrt{E}n}}{T_\omega^{(1,n)}(E)^\ast}
+ \frac{e^{-i\sqrt{E}n}}{T_\omega^{(1,n)}(E)}\right],\\
\left[\widetilde{\phi}_\omega(n;E)\right]_{12} &=& \frac{1}{2i\sqrt{E}}
\left[\frac{L_\omega^{(1,n)}(E)^\ast}{T_\omega^{(1,n)}(E)^\ast}
e^{-i\sqrt{E}(n+1)}-
\frac{L_\omega^{(1,n)}(E)}{T_\omega^{(1,n)}(E)}e^{i\sqrt{E}(n+1)}\right]\\
&& +\frac{1}{2i\sqrt{E}}\left[\frac{e^{i\sqrt{E}n}}{T_\omega^{(1,n)}(E)^\ast}
- \frac{e^{-i\sqrt{E}n}}{T_\omega^{(1,n)}(E)}\right],\\
\left[\widetilde{\phi}_\omega(n;E)\right]_{21} &=& -\frac{i\sqrt{E}}{2}
\left[\frac{L_\omega^{(1,n)}(E)^\ast}{T_\omega^{(1,n)}(E)^\ast}
e^{-i\sqrt{E}(n+1)}-
\frac{L_\omega^{(1,n)}(E)}{T_\omega^{(1,n)}(E)}e^{i\sqrt{E}(n+1)}\right]\\
&& +\frac{i\sqrt{E}}{2}\left[\frac{e^{i\sqrt{E}n}}{T_\omega^{(1,n)}(E)^\ast}
- \frac{e^{-i\sqrt{E}n}}{T_\omega^{(1,n)}(E)}\right],\\
\left[\widetilde{\phi}_\omega(n;E)\right]_{22} &=& -\frac{1}{2}
\left[\frac{L_\omega^{(1,n)}(E)^\ast}{T_\omega^{(1,n)}(E)^\ast}
e^{-i\sqrt{E}(n+1)}+
\frac{L_\omega^{(1,n)}(E)}{T_\omega^{(1,n)}(E)}e^{i\sqrt{E}(n+1)}\right]\\
&& +\frac{1}{2}\left[\frac{e^{i\sqrt{E}n}}{T_\omega^{(1,n)}(E)^\ast}
+ \frac{e^{-i\sqrt{E}n}}{T_\omega^{(1,n)}(E)}\right].
\end{eqnarray*}
Now using the relation (see (\ref{s.param}))
\begin{displaymath}
L_\omega^{(1,n)}(E)=i\sqrt{1-|T_\omega^{(1,n)}(E)|^2}\
e^{i\delta_\omega^{(1,n)}(E)-i\theta_\omega^{(1,n)}(E)},
\end{displaymath}
we calculate the Hilbert-Schmidt norm of $\widetilde{\phi}_\omega(n;E)$.
After some simple transformations we get that
\begin{eqnarray}\label{long.1}
\|\widetilde{\phi}_\omega(n;E) \|_{\mathcal{J}_2}^2=|T_\omega^{(1,n)}(E)|^{-2}
\Big\{2(1-|T_\omega^{(1,n)}(E)|^2)\cos^2\theta_1+2\cos^2\theta_2\nonumber\\
+\frac{1}{E}\left(\sqrt{1-|T_\omega^{(1,n)}(E)|^2}\sin\theta_1
-\sin\theta_2 \right)^2\nonumber\\
+E\left(\sqrt{1-|T_\omega^{(1,n)}(E)|^2}\sin\theta_1 +\sin\theta_2
\right)^2\Big\}.
\end{eqnarray}
Here for brevity we have introduced the notations
\begin{eqnarray*}
\theta_1 &=& \sqrt{E}(n+1)-\theta_\omega^{(1,n)}(E)+\pi/2,\\
\theta_2 &=& \sqrt{E}n+\delta_\omega^{(1,n)}(E).
\end{eqnarray*}
Obviously, the expression in the braces on the r.h.s. of (\ref{long.1}) is bounded from
above for every $E>0$ uniformly in $n\in\N$ by $4(1+E+1/E)$. In Appendix A we show that
this expression is also bounded from below by the positive constant
\begin{displaymath}
C(E)=\frac{4E}{1+E^2}
\end{displaymath}
uniformly in $n\in\N$. Thus,
\begin{eqnarray}\label{aux.4}
\gamma(E)=\lim_{n\rightarrow\infty}\frac{1}{n}
\log\|\widetilde{\phi}_\omega(n;E)\|\nonumber\\
=-\lim_{n\rightarrow\infty}\frac{1}{n}\log|T_\omega^{(1,n)}(E)|.
\end{eqnarray}
Now consider
\begin{eqnarray*}
\gamma_T(E) &=& -\lim_{n,m\rightarrow\infty}\frac{\log|T_\omega^{(-m,n)}(E)|}
{n+m+1}\\
&=& -\lim_{k\rightarrow\infty}\lim_{N\rightarrow\infty}
\frac{\log|T_\omega^{(1-k,N-k)}(E)|}{N}\\
&=& -\lim_{k\rightarrow\infty}\lim_{N\rightarrow\infty}
\frac{\log|T_{T_k\omega}^{(1,N)}(E)|}{N},
\end{eqnarray*}
where $N=n+m+1$ and $k=1+m$. By (\ref{aux.4}) for $\P$-almost all
$\omega$
\begin{displaymath}
-\lim_{N\rightarrow\infty}\frac{\log|T_\omega^{(1,N)}(E)|}{N}=
\gamma(E)
\end{displaymath}
independently of $k$. $\blacksquare$

\vspace{0.25in}

Now we recall that $\arg T_\omega^{(-n,m)}(E) =-\pi\xi^{(-n,m)}(E;\omega)$. Then by
(\ref{limit}) and Theorems 5.1, 5.3 we have

\vspace{0.25in}

\bf Corollary 5.4 \it For every $E>0$ and $\P$-almost all $\omega$
\begin{equation}\label{Col.4}
\lim_{m,n\rightarrow\infty}\frac{\log T_\omega^{(-n,m)}(E)}{m+n+1}=
-\gamma(E)-i\pi\xi(E).
\end{equation}
\rm

\vspace{0.25in}

We note that Corollary 5.4 can be reformulated in such a way that it permits a
generalization to the higher-dimensional case. We recall that (see \cite{Newton})
\begin{equation}\label{relation}
T_\omega^{(-n,m)}(E)^{-1}=\det\left(I+{V_\omega^{(-n,m)}}^{1/2}
R_0(E+i0){|V_\omega^{(-n,m)}|}^{1/2} \right),
\end{equation}
where
\begin{displaymath}
V_\omega^{(-n,m)}(x)=\sum_{j=-n}^m \alpha_j(\omega)f(x-j).
\end{displaymath}
Now we can rewrite (\ref{Col.4})  as follows
\begin{eqnarray*}
\lim_{m,n\rightarrow\infty}\frac{1}{m+n+1}\log\det
\left(I+{V_\omega^{(-n,m)}}^{1/2}R_0(E+i0){|V_\omega^{(-n,m)}|}^{1/2} \right)\\=
\gamma(E)+i\pi\xi(E).
\end{eqnarray*}
In this form (\ref{Col.4}) can now be generalized to the higher-dimensional case, thus
defining $\gamma(E)$, which is something like the multi-dimensional Lyapunov exponent (for
details  see \cite{KoSch3}). Also note that the above determinant equals the determinant
of the Jost matrix \cite{Newton}.

We cannot use (\ref{Col.4}) to calculate $\gamma(E)$ for $E<0$. However, this can be done
by means of analytic continuation (see Section 6).

\vspace{0.25in}

We turn to the claim $\gamma(E)>0$ for almost all $E>0$. We start with some preparations.
Let us denote
\begin{displaymath}
\Lambda^{(-n,m)}(E;\omega)=\left(\begin{array}{lr}
                               \frac{1}{T_\omega^{(-n,m)}(E)} & -\frac{R_\omega^{(-n,m)}(E)}
                               {T_\omega^{(-n,m)}(E)}\\
                               \frac{L_\omega^{(-n,m)}(E)}{T_\omega^{(-n,m)}(E)} &
                               \frac{1}{T_\omega^{(-n,m)}(E)^\ast}
                               \end{array} \right).
\end{displaymath}
By the identity
\begin{displaymath}
\|\Lambda^{(-n,m)}(E;\omega)\|_{\mathcal{J}_2}^2=
\frac{2+|R_\omega^{(-n,m)}|^2+|L_\omega^{(-n,m)}(E)|^2}{|T_\omega^{(-n,m)}(E)|^2}=
\frac{4-|T_\omega^{(-n,m)}(E)|^2}{|T_\omega^{(-n,m)}(E)|^2},
\end{displaymath}
and by Theorem 5.3 one has that for every $E>0$
\begin{equation}\label{gamma.E.1}
\gamma(E)=\lim_{m,n\rightarrow\infty}\frac{1}{n+m+1}\log\|\Lambda^{(-n,m)}(E;\omega) \|
\end{equation}
almost surely. Also (\ref{gamma.E.1}) follows directly from the definition of the Lyapunov
exponent (\ref{Lyap.def}) and Theorem 2.4.

Let $H_\alpha:=H_0+\alpha f$ for some $\alpha\in\R$. The corresponding elements of the
scattering matrix at energy $E>0$ we denote by $T_\alpha(E)$, $R_\alpha(E)$, and
$L_\alpha(E)$. Let
\begin{displaymath}
\Lambda_\alpha(E)=\left(\begin{array}{lr}
                   \frac{1}{T_\alpha (E)} & -\frac{R_\alpha (E)}{T_\alpha (E)}\\
                   \frac{L_\alpha (E)}{T_\alpha (E)} & \frac{1}{T_\alpha (E)^\ast}
                   \end{array}\right),
\end{displaymath}
and
\begin{displaymath}
\Lambda_j(E;\omega)=\left(\begin{array}{lr}
                   \frac{1}{T_{\alpha_j(\omega)} (E)} & -\frac{R_{\alpha_j(\omega)} (E)}
                   {T_{\alpha_j(\omega)} (E)}e^{-2i\sqrt{E}j}\\
                   \frac{L_{\alpha_j(\omega)} (E)}{T_{\alpha_j(\omega)} (E)}e^{2i\sqrt{E}j} &
                   \frac{1}{T_{\alpha_j(\omega)} (E)^\ast}
                   \end{array}\right).
\end{displaymath}
Obviously,
\begin{displaymath}
\Lambda_j(E;\omega)=U_E^j\Lambda_{\alpha_j(\omega)}(E) U_E^{-j},
\end{displaymath}
where
\begin{displaymath}
U_E=\left(\begin{array}{cc}
           e^{-i\sqrt{E}} & 0 \\
           0 & e^{i\sqrt{E}}
           \end{array}\right).
\end{displaymath}
From the Aktosun factorization formula (\ref{Aktosun.form}) it follows that
\begin{eqnarray}\label{Lambda.U}
\Lambda^{(-n,m)}(E;\omega)=\prod_{j=-n}^m \Lambda_j(E;\omega)=
\prod_{j=-n}^m U_E^j\Lambda_{\alpha_j(\omega)}(E)U_E^{-j}\nonumber\\
=U_E^{-n-1/2}\prod_{j=-n}^m U_E^{1/2}\Lambda_{\alpha_j(\omega)}(E)U_E^{1/2}\cdot
U_E^{-m-1/2}.
\end{eqnarray}
Since $U_E$ is unitary one obtains
\begin{equation}\label{gamma.E.2}
\|\Lambda^{(-n,m)}(E;\omega)\|=\left\|\prod_{j=-n}^m\widetilde{\Lambda}_{\alpha_j(\omega)}(E)
 \right\|
\end{equation}
with
\begin{displaymath}
\widetilde{\Lambda}_\alpha(E)=U_E^{1/2}\Lambda_\alpha(E) U_E^{1/2}=
\left(\begin{array}{cc}
       \frac{e^{-i\sqrt{E}}}{T_\alpha(E)} & -\frac{R_\alpha (E)}{T_\alpha (E)}\\
       \frac{L_\alpha (E)}{T_\alpha (E)} & \frac{e^{i\sqrt{E}}}{T_\alpha(E)^\ast}
       \end{array} \right).
\end{displaymath}
Since the $\alpha_j(\omega)$ form a sequence of random i.i.d. variables, for every $E>0$
the sequence $\widetilde{\Lambda}_{\alpha_j(\omega)}(E)$ is a sequence of random i.i.d.
$\SL(2;\C)$-valued variables with corresponding distribution $\widetilde{\varphi}_E$.
Recall that the distribution density $\varphi$ is supposed to be continuous with compact
support.

The matrices $\widetilde{\Lambda}_\alpha(E)$ have the form
\begin{displaymath}
\left(\begin{array}{ll}
      a  &  b\\
      b^\ast & a^\ast
      \end{array} \right).
\end{displaymath}
The closed subgroup of all matrices from $\SL(2;\C)$ having this form (which we denote by
$\SL_\R(2;\C)$) is isomorphic to $\SL(2;\R)$. Indeed,
\begin{displaymath}
\left(\begin{array}{ll}
      a  &  b\\
      b^\ast & a^\ast
      \end{array} \right)=Q\left(\begin{array}{cc}
      {\Re}a+{\Re}b & -{\Im}a+{\Im}b \\
      {\Im}a+{\Im}b & {\Re}a-{\Re}b
      \end{array} \right) Q^{-1},
\end{displaymath}
with
\begin{displaymath}
Q=\frac{1}{2}\left(\begin{array}{cc}
      1-i  &  1+i\\
      1-i & -1-i
      \end{array} \right).
\end{displaymath}

Let $\xi_\alpha(E)$ be the spectral shift function for the pair of Hamiltonians
($H_\alpha$, $H_0$) such that $T_\alpha(E)=|T_\alpha(E)|e^{-i\pi\xi_\alpha(E)}$ for $E>0$.
It is well known (see e.g. \cite{KiKoSi}) that $T_\alpha(E)$, $R_\alpha(E)$, and
$L_\alpha(E)$ at fixed energy $E>0$ are real analytic functions of $\alpha\in\R$. Since
$f$ has compact support, $T_\alpha(E)$, $R_\alpha(E)$, and $L_\alpha(E)$ at fixed
$\alpha\in\R$ are real analytic with respect to $E>0$ \cite{Deift:Trubowitz}. Moreover,
they are jointly real analytic in $\alpha\in\R$ and $E>0$. Since $T_\alpha(E)\neq 0$ for
all $E>0$, $|T_\alpha(E)|$ and $\xi_\alpha(E)$ are also jointly real analytic in
$\alpha\in\R$ and $E>0$.

We recall that potentials with compact support cannot be reflectionless, i.e.
$R_\alpha(E)=0$ for all $E>0$ implies $\alpha=0$ \cite{Deift:Trubowitz}.

By real analyticity the set
\begin{displaymath}
S_\alpha=\left\{E>0: R_\alpha(E)=0 \right\}
\end{displaymath}
for every fixed $\alpha\neq 0$ is discrete or empty. Let
\begin{displaymath}
S=\left\{E>0: R_\alpha(E)=0\ \textrm{for all}\ \alpha\in\supp\varphi \right\}.
\end{displaymath}
By the assumption that $\supp\varphi$ has a positive Lebesgue measure and by real
analyticity of $R_\alpha(E)$, the condition $R_\alpha(E)=0$ for all
$\alpha\in\supp\varphi$ or even for $\alpha$ in a subset of $\supp\varphi$ of positive
measure implies that $R_\alpha(E)=0$ for all $\alpha\in\R$. Obviously,
$S=\cap_{\alpha\in\R}S_\alpha$ and therefore is also discrete or even empty.

\vspace{0.25in}

\textit{Remark:} One can easily show that a necessary (but not sufficient) condition for
$E\in S$ is
\begin{displaymath}
\int e^{2i\sqrt{E}x}f(x)dx=0.
\end{displaymath}
We know, however, no example of a potential with $S\neq\emptyset$ and expect that actually
$S=\emptyset$, since, intuitively, it is clear that for the potential $f\geq 0$ at hand
$T_\alpha(E)\rightarrow 0$ as $\alpha\rightarrow\infty$.

\vspace{0.25in}

Now for every fixed $E>0$, $E\notin S$ we define the functions
\begin{displaymath}
F_E^{(\pm)}(\alpha)=\frac{-i\sin(\sqrt{E}-\pi\xi_\alpha(E))
\pm\sqrt{\cos^2(\sqrt{E}-\pi\xi_\alpha(E))-|T_\alpha(E)|^2}}
{R_\alpha(E)e^{i\pi\xi_\alpha(E)}}
\end{displaymath}
Let
\begin{displaymath}
S^{(\pm)}=\left\{E>0,\ E\notin S:\ F_E^{(\pm)}(\alpha)\ \textrm{does not depend on}\
\alpha\in\supp\varphi
\right\}.
\end{displaymath}

If for some $E>0$ one of the functions $F_E^{(\pm)}=C^{(\pm)}=\const$ for all
$\alpha\in\supp\varphi$, then
\begin{displaymath}
\sqrt{\cos^2(\sqrt{E}-\pi\xi_\alpha(E))-|T_\alpha(E)|^2}=\pm iC^{(\pm)}R_\alpha(E)
e^{i\pi\xi_\alpha(E)}\pm i\sin(\sqrt{E}-\pi\xi_\alpha(E))
\end{displaymath}
is real analytic with respect to $\alpha\in\R$ and hence $F_E^{(\pm)}(\alpha)=C^{(\pm)}$
for all $\alpha\in\R$. Also, it follows that the zeros of
$\cos^2(\sqrt{E}-\pi\xi_\alpha(E))-|T_\alpha(E)|^2$ (if any) are all of even order. Thus,
we have that if $E>0$ belongs to one of the sets $S^{(\pm)}$, then either
\begin{equation}\label{case.one}
\cos^2(\sqrt{E}-\pi\xi_\alpha(E))\leq |T_\alpha(E)|^2,
\end{equation}
or
\begin{equation}\label{case.two}
\cos^2(\sqrt{E}-\pi\xi_\alpha(E))\geq |T_\alpha(E)|^2,
\end{equation}
for all $\alpha\in\R$. Taking $\alpha=0$ in (\ref{case.two}) and using
$\xi_{\alpha=0}(E)=0$ and $T_{\alpha=0}(E)=1$ gives that $E=(\pi n)^2$ with some $n\in\N$.

Now calculating the limit $\alpha\rightarrow 0$ of $F_E^{(\pm)}(\alpha)$ and taking into
account that $R_\alpha(E)$ is not identically zero, we obtain that for $E\in S^{(\pm)}$
\begin{displaymath}
\sin\sqrt{E}\mp|\sin\sqrt{E}|=0,
\end{displaymath}
respectively. Thus,
\begin{eqnarray}\label{set}
S^{(+)}&\subseteq&\bigcup_{n=0}^\infty[4\pi^2 n^2, \pi^2(2n+1)^2], \\ && \\
\nonumber S^{(-)}
&\subseteq&\bigcup_{n=0}^\infty[\pi^2 (2n+1)^2, 4\pi^2(n+1)^2].\nonumber
\end{eqnarray}
Now let us suppose that $E$ belongs to the interior of the sets on the r.h.s. of
(\ref{set}). This implies that $\sin\sqrt{E}>0$ if $E\in S^{(+)}$ and $\sin\sqrt{E}<0$ if
$E\in S^{(-)}$. Then using
$R_\alpha(E)=i|R_\alpha(E)|\exp\{i\theta_\alpha(E)-i\pi\xi_\alpha(E)
\}$, we calculate $F_E^{(\pm)}(\alpha)$ (with the corresponding choice of sign) for small
$\alpha$:
\begin{eqnarray*}
F_E^{(\pm)}(\alpha)&=&\frac{-\sin(\sqrt{E}-\pi\xi_\alpha(E))\pm
|\sin(\sqrt{E}-\pi\xi_\alpha(E))|\sqrt{1-\frac{|R_\alpha(E)|^2}
{\sin^2(\sqrt{E}-\pi\xi_\alpha(E))}}}{|R_\alpha(E)|e^{i\theta_\alpha(E)}}\\
&=&\mp\frac{|R_\alpha(E)|}{2|\sin(\sqrt{E}-\pi\xi_\alpha(E))|}e^{-i\theta_\alpha(E)}
+O(|R_\alpha(E)|^3).
\end{eqnarray*}
Thus $F_E^{(\pm)}(\alpha=0)=0$ and therefore $F_E^{(\pm)}(\alpha)=0$ for all
$\alpha\in\R$. This implies that
\begin{displaymath}
-i\sin(\sqrt{E}-\pi\xi_\alpha(E))\pm\sqrt{|R_\alpha(E)|^2-\sin^2(\sqrt{E}-\pi\xi_\alpha(E))}=0
\end{displaymath}
for all $\alpha\in\R$. Hence $R_\alpha(E)=0$ for all $\alpha\in\R$, which contradicts the
assumption $E\notin S$.

For the sake of convenience we summarize some of the established properties of the sets
$S^{(\pm)}$:

\vspace{0.25in}

\textbf{Lemma 5.5} \it The sets $S^{(\pm)}$ are at most discrete. More precisely,
\begin{displaymath}
S^{(\pm)}\subseteq \{(\pi n)^2,\ n\in\N \}.
\end{displaymath}
If $E\in S^{(\pm)}$ then either (\ref{case.one}) or (\ref{case.two}) holds. \rm

\vspace{0.25in}

Now we define
\begin{displaymath}
\widetilde{S}=\left(S^{(+)}\cup S^{(-)}\right)\cap\{E>0:
\cos^2(\sqrt{E}-\pi\xi_\alpha(E))\leq |T_\alpha(E)|^2\ \textrm{for all}\
\alpha\in\R\},
\end{displaymath}
which is at most discrete.

\vspace{0.25in}

\textbf{Theorem 5.6} \it For almost all $E>0$ the upper Lyapunov exponent
$\gamma(E)>0$ almost surely. More precisely $\gamma(E)$ vanishes for $E\in
S\cup\widetilde{S}$ and almost surely nowhere else.
\rm

\vspace{0.25in}

\textit{Proof}. We split the proof in several steps and start with the case $E\in S$.
Calculating the eigenvalues of
$\widetilde{\Lambda}_\alpha(E)\widetilde{\Lambda}_\alpha(E)^\ast$ one easily finds that
\begin{displaymath}
\|\widetilde{\Lambda}_\alpha(E)\|^2=\frac{2-|T_\alpha(E)|^2}{|T_\alpha(E)|^2}+
\sqrt{\left( \frac{2-|T_\alpha(E)|^2}{|T_\alpha(E)|^2}\right)^2-1}\geq 1,
\end{displaymath}
where the norm is understood in the operator sense. Obviously,
$\|\widetilde{\Lambda}_\alpha(E)\|=1$ iff $R_\alpha(E)=0$. Therefore, if $E\in S$ then
\begin{displaymath}
\log\left\|\prod_{j=-n}^m \widetilde{\Lambda}_{\alpha_j(\omega)}(E) \right\|\leq
\log\prod_{j=-n}^m\|\widetilde{\Lambda}_{\alpha_j(\omega)}(E) \|=0,
\end{displaymath}
and hence $\gamma(E)=0$.

To proceed further we define the family of auxiliary periodic Hamiltonians
\begin{equation}\label{Hill}
H^{(\alpha)}=H_0+\alpha\sum_{j\in\Z}f(\cdot-j).
\end{equation}
The spectrum of every $H^{(\alpha)}$ is purely absolute continuous and has a band
structure. We recall (see e.g. \cite{Magnus:Winkler}) that the discriminant
$\Delta_\alpha(E)$ of (\ref{Hill}) is defined by $\Delta_\alpha(E)=u_1(1)+ u'_2(1)$, where
$u_1(x)$ and $u_2(x)$ are solutions of $H^{(\alpha)}u_i=E u_i$ with the initial data
$u_1(0)=u'_2(0)=1$, $u'_1(0)=u_2(0)=0$. $\Delta_\alpha(z)$ is an entire function of
$z\in\C$. The real solutions of the inequality $|\Delta_\alpha(E)|>2$ determine the gaps
in the spectrum of $H^{(\alpha)}$, whereas $|\Delta_\alpha(E)|<2$ implies that $E$ belongs
to the spectrum.

We say that
\textit{$E$ is in a gap of $H^{(\alpha)}$} for some $\alpha\in\R$ if there is $\delta>0$
such that $(E-\delta,E+\delta)\cap\sigma(H^{(\alpha)})=\emptyset$.

Keller \cite{Keller} proved that for $E>0$,
\begin{displaymath}
\Delta_\alpha(E)=\frac{2\cos(\sqrt{E}-\pi\xi_\alpha(E))}{|T_\alpha(E)|}.
\end{displaymath}
Thus, $E>0$ is in a gap of $H^{(\alpha)}$ iff $\cos^2(\sqrt{E}-
\pi\xi_\alpha(E))>|T_\alpha(E)|^2$.

\vspace{0.25in}

\textit{Remark:} We sketch another proof of this fact based on the Ishii-Pastur-Kotani
theorem \cite{Kotani}. If for some $E>0$
$\cos^2(\sqrt{E}-\pi\xi_\alpha(E))>|T_\alpha(E)|^2$, then one of the eigenvalues of
$\widetilde{\Lambda}_\alpha(E)$,
\begin{equation}\label{eigenvalues}
\lambda_\pm(\alpha)=\frac{\cos(\sqrt{E}-\pi\xi_\alpha(E))}{|T_\alpha(E)|}\mp
\sqrt{\frac{\cos^2(\sqrt{E}-\pi\xi_\alpha(E))}{|T_\alpha(E)|^2}-1},
\end{equation}
is strictly larger than 1 in absolute value. Therefore, for the Lyapunov exponent
corresponding to $H^{(\alpha)}$ we have
\begin{eqnarray*}
\gamma_\alpha(E) &=& \lim_{m,n\rightarrow\infty}\frac{1}{n+m+1}\log\|\widetilde{\Lambda}_\alpha
(E)^{n+m+1}\|\\ &\geq & \lim_{m,n\rightarrow\infty}\frac{1}{n+m+1}\log
\max_\pm|\lambda_\pm(\alpha)|^{n+m+1}\\
&=& \log\max_\pm|\lambda_\pm(\alpha)|>0.
\end{eqnarray*}
Since the spectrum of $H^{(\alpha)}$ is absolutely continuous, by the Ishii-Pastur-Kotani
theorem $E$ lies in a gap. Conversely, if
$\cos^2(\sqrt{E}-\pi\xi_\alpha(E))\leq|T_\alpha(E)|^2$, then both eigenvalues
(\ref{eigenvalues}) lie on the unit circle and the Lyapunov exponent vanishes,
\begin{displaymath}
\gamma_\alpha(E)=\lim_{m,n\rightarrow\infty}\frac{1}{n+m+1}\log\|\widetilde{\Lambda}_\alpha
(E)^{n+m+1}\|=0.
\end{displaymath}
Hence $E$ belongs to the spectrum.

\vspace{0.25in}

We proceed with the proof of Theorem 5.6. Let us suppose that $E\in\R\setminus S$. Let
$G_E\subset\SL(2;\C)$ be the smallest closed subgroup which contains the support of
$\widetilde{\varphi}$ (recall that $\widetilde{\varphi}$ is the image of $\varphi$ under
the map $\alpha\mapsto\widetilde{\Lambda}_\alpha(E)$). To establish that $\gamma(E)>0$ for
almost all $E>0$ almost surely we will also use the sufficient conditions of the
Furstenberg theorem \cite[Theorem A.II.4.1, Proposition A.II.4.3]{BougerolLacroix} (in the
complex form), i.e. for almost all $E>0$ the group $G_E$ is not compact and for any
$\mathbf{x}\in
\C\P^1$ the set $\{G_E\cdot\mathbf{x}\}\subset\C\P^1$ has more than two elements.

We start with the second condition in the Furstenberg theorem.
First we calculate the
(non normalized) eigenvectors of $\widetilde{\Lambda}_\alpha(E)$,
\begin{displaymath}
g_\alpha^{(\pm)}(E)=\frac{1}{|T_\alpha(E)|}\left(\begin{array}{c}
                     R_\alpha(E)e^{i\pi\xi_\alpha(E)}\\
                     -i\sin(\sqrt{E}-\pi\xi_\alpha(E))\pm\sqrt{\cos^2(\sqrt{E}-\pi\xi_\alpha(E))
                     -|T_\alpha(E)|^2}
                     \end{array}\right).
\end{displaymath}
In the case $\cos^2(\sqrt{E}-\pi\xi_\alpha(E))=|T_\alpha(E)|^2$ both eigenvectors coincide
and the corresponding generalized eigenvector (i.e. the solution of
$(\widetilde{\Lambda}_\alpha(E)-1)\widetilde{g}_\alpha(E)=g^{(\pm)}_\alpha(E)$) is
$(0,-1)^T$.

By means of the bijection $p:\C\P^1\rightarrow\C\cup\{\infty\}$,
\begin{displaymath}
p\left(\begin{array}{c}
       x_1 \\
       x_2\end{array}\right)=\left\{\begin{array}{ll}
                             \frac{x_2}{x_1}, & x_1\neq 0,\\
                             \infty, & x_1=0,\end{array}\right.
\end{displaymath}
we can identify the complex projective line $\C\P^1$ and $\C\cup\{\infty\}$. Obviously,
\begin{displaymath}
p(g_\alpha^{(\pm)}(E))=F_E^{(\pm)}(\alpha).
\end{displaymath}
For fixed $E$ the reflection amplitude $R_\alpha(E)$ as a function of
$\alpha\in\supp\varphi$ has at most a discrete set of zeros.

Actually (see \cite[Problem 6.4]{BougerolLacroix}) for any subgroup
$G\subseteq\SL_\R(2;\C)$ one has the alternatives:
\newcounter{punkt}
\begin{list}{(\roman{punkt})}{\usecounter{punkt}}
\item $G$ is finite,
\item there is $Q\in\GL(2;\C)$ such that
\begin{displaymath}
Q^{-1}GQ\subseteq\left\{\left(\begin{array}{cc}
                      a & 0 \\
                      0 & a^{-1}\end{array}\right),\ a\in\C\setminus\{0\}\right\}\cup
\left\{\left(\begin{array}{cc}
                      0 & b^{-1} \\
                      -b & 0\end{array}\right),\ b\in\C\setminus\{0\}\right\},
\end{displaymath}
\item there is $Q\in\GL(2;\C)$ such that
\begin{displaymath}
Q^{-1}GQ\subseteq\left\{\left(\begin{array}{cc}
                      a & b \\
                      0 & a^{-1}\end{array}\right),\ a\in\C\setminus\{0\},\ b\in\C\right\},
\end{displaymath}
\item for any $\mathbf{x}\in\C\P^1$ the set $\{G\cdot\mathbf{x}\}\subset\C\P^1$ has
more than two elements.
\end{list}

Let us first suppose that $G=G_E$ is finite. Since $\widetilde{\Lambda}_\alpha(E)$ is real
analytic with respect to $\alpha$ this implies that $\widetilde{\Lambda}_\alpha(E)$ is
constant for all $\alpha\in\R$ and hence $|T_\alpha(E)|$ does not depend on $\alpha$.
Therefore $|T_\alpha(E)|=|T_{\alpha=0}(E)|=1$ and thus $E\in S$.

Consider the case (ii) for $G=G_E$. Let us suppose that for some $Q$ the matrices
$Q^{-1}\widetilde{\Lambda}_\alpha(E)Q$ are diagonal for all $\alpha$ in a subset of
$\supp\varphi$ of positive measure. The existence of such $Q$ implies that the
eigenvectors of all $\widetilde{\Lambda}_\alpha(E)$ are constant as elements of $\C\P^1$
and thus
\begin{equation}\label{Stern.Stern}
p(g_\alpha^{(\pm)}(E))=C^{(\pm)}=\const
\end{equation}
with $C^{(+)}\neq C^{(-)}$ for all such $\alpha$. But then these relations hold for all
$\alpha\in\R$ by real analyticity. Thus $E\in S^{(+)}\cap S^{(-)}$. Conversely, if $E\in
S^{(+)}\cap S^{(+)}$ (such that $\cos^2\sqrt{E}= 1$) and if in addition
$\cos^2(\sqrt{E}-\pi\xi_\alpha(E))\neq|T_\alpha(E)|^2$ for all $\alpha$ in a subset of
$\supp\varphi$ of positive measure, then
\begin{displaymath}
Q=\left(\begin{array}{cc}
        1 & 1 \\
        p(g_\alpha^{(+)}(E)) & p(g_\alpha^{(-)}(E))
        \end{array} \right)\in\GL(2;\C)
\end{displaymath}
does not depend on $\alpha$ in this set and $Q^{-1}\widetilde{\Lambda}_\alpha(E)Q$ is
diagonal for all these $\alpha$. Hence for this choice of $Q$ and by real analyticity
$Q^{-1}\widetilde{\Lambda}_\alpha Q$ is diagonal for all $\alpha\in\R$. By the previous
discussion (see Lemma 5.5) in case $\cos^2(\sqrt{E}-\pi\xi_\alpha(E))\leq |T_\alpha(E)|^2$
with equality for $\alpha$ only on a set of measure zero the eigenvalues of
$\widetilde{\Lambda}_{\alpha_j(\omega)}(E)$ lie on the unit circle. Therefore, since
\begin{displaymath}
\prod_{j=-n}^m\widetilde{\Lambda}_{\alpha_j(\omega)}(E)=Q\left(
\begin{array}{cc}
\prod_{j=-n}^m\lambda_+(\alpha_j(\omega)) & 0 \\
0 & \prod_{j=-n}^m\lambda_-(\alpha_j(\omega))\end{array} \right)Q^{-1},
\end{displaymath}
we have $\|\prod_{j=-n}^m\widetilde{\Lambda}_{\alpha_j(\omega)}(E)\|=1$. Thus
$\gamma(E)=0$.

Consider the opposite case when $\cos^2(\sqrt{E}-\pi\xi_\alpha(E))\geq|T_\alpha(E)|^2$
with equality for $\alpha$ only on a set of measure zero. In this case one of the
eigenvalues $\lambda_\pm(\alpha)$ (say $\lambda_+(\alpha)$) for almost all
$\alpha\in\supp\varphi$ is larger than 1 in absolute value. Since $E\in S^{(+)}\cap
S^{(-)}$ the eigenvalues of $\prod_{j=-n}^m\widetilde{\Lambda}_{\alpha_j(\omega)}(E)$ are
given by $\prod_{j=-n}^m\lambda_+(\alpha_j(\omega))$ and
$\prod_{j=-n}^m\lambda_-(\alpha_j(\omega))$ and thus
\begin{eqnarray*}
\gamma(E) &\geq& \lim_{n,m\rightarrow\infty}\frac{1}{m+n+1}\sum_{j=-n}^m
\log|\lambda_+(\alpha_j(\omega))|\\
&=& \E\left\{\log|\lambda_+(\alpha_j(\omega))|\right\}>0
\end{eqnarray*}
almost surely. Here we used the fact that $\{\alpha_j(\omega)\}_{j\in\Z}$ is metrically
transitive and the Birkhoff-Khintchin theorem.

It remains to consider the case when $E\in S^{(+)}\cap S^{(-)}$ and
$\cos^2(\sqrt{E}-\pi\xi_\alpha(E))=|T_\alpha(E)|^2$ for almost all $\alpha$ in
$\supp\varphi$. By real analyticity this relation holds for all $\alpha\in\R$. But then
$\lambda_+(\alpha)=\lambda_-(\alpha)=\pm 1$ and hence
$Q^{-1}\widetilde{\Lambda}_\alpha(E)Q=
\widetilde{\Lambda}_\alpha(E)=\pm I$ for all $\alpha\in\R$. Evaluating at $\alpha=0$ gives
$T_\alpha(E)=1$ for all $\alpha$ and thus $E\in S$. We note that from the real analyticity
with respect to $\alpha$ it follows that if there exists $Q\in\GL(2;\C)$ such that
$Q^{-1}\widetilde{\Lambda}_\alpha(E)Q$ is diagonal for $\alpha$ on a set of positive
measure, then $Q^{-1}\widetilde{\Lambda}_\alpha(E)Q$ is diagonal for all $\alpha\in\R$.

To conclude the discussion of case (ii) for $G=G_E$ suppose now that there is
$Q\in\GL(2;\C)$ such that for almost all $\alpha\in\supp\varphi$
\begin{equation}\label{Q.off}
Q^{-1}\widetilde{\Lambda}_\alpha(E) Q=\left(\begin{array}{cc}
                                       0 & b(\alpha)^{-1} \\
                                       -b(\alpha) & 0 \end{array}\right)
\end{equation}
holds for some $b(\alpha)\in\C\setminus\{0\}$. By real analyticity in $\alpha$ of
$\widetilde{\Lambda}_\alpha(E)$ such a relation holds for all $\alpha\in\R$. Taking traces
gives
\begin{displaymath}
\cos(\sqrt{E}-\pi\xi_\alpha(E))=0
\end{displaymath}
for all $\alpha\in\R$. Again by real analyticity this implies
$\xi_\alpha(E)=\const=\xi_{\alpha=0}(E)=0$. In particular, (\ref{Q.off}) cannot hold if
$\cos\sqrt{E}\neq 0$. In case $\cos\sqrt{E}=0$ observe that for $b\neq 0$
\begin{displaymath}
\left(\begin{array}{cc}
0 & b^{-1} \\
-b & 0 \end{array} \right)=\mathcal{R}_b^{-1}\left(\begin{array}{cc}
i & 0 \\ 0 & -i \end{array} \right)\mathcal{R}_b,\ \mathcal{R}_b=\left(\begin{array}{cc} b
& -i \\ b & i \end{array} \right),
\end{displaymath}
and
\begin{displaymath}
Q_\alpha^{-1}\widetilde{\Lambda}_\alpha(E)Q_\alpha=\left(\begin{array}{cc} i & 0 \\ 0 & -i
\end{array}\right)
\end{displaymath}
with
\begin{displaymath}
Q_\alpha=\left(\begin{array}{cc} R_\alpha(E) & R_\alpha(E)
\\ -i\sin\sqrt{E}-i|T_\alpha(E)| &
-i\sin\sqrt{E}+i|T_\alpha(E)|
\end{array}
\right).
\end{displaymath}
Hence any matrix which diagonalizes $\widetilde{\Lambda}_\alpha(E)$ is necessary of the
form $Q_\alpha\left(\begin{array}{cc}a & 0\\ 0 & d\end{array} \right)$ with $ad\neq 0$.
This implies that
\begin{equation}\label{bez}
Q=\left(\begin{array}{cc} q_1 & q_2 \\ q_3 & q_4\end{array}\right)= Q_\alpha
\left(\begin{array}{cc} a(\alpha) & 0 \\ 0 & d(\alpha)\end{array} \right)
\mathcal{R}_{b(\alpha)}
\end{equation}
for almost all $\alpha\in\supp\varphi$, where $a(\alpha)$ and $d(\alpha)$ are suitable
nonvanishing functions. Writing (\ref{bez}) explicitly gives
\begin{eqnarray*}
q_1 &=& b(\alpha)(a(\alpha)+d(\alpha))R_\alpha(E),\\ q_2 &=&
-i(a(\alpha)-d(\alpha))R_\alpha(E),\\
q_3 &=&
-ib(\alpha)(a(\alpha)+d(\alpha))\sin\sqrt{E}-ib(\alpha)(a(\alpha)-d(\alpha))|T_\alpha(E)|,\\
q_4 &=& -(a(\alpha)-d(\alpha))\sin\sqrt{E}-(a(\alpha)+d(\alpha))|T_\alpha(E)|.
\end{eqnarray*}
From this it follows that for all $\alpha\in\supp\varphi$
\begin{eqnarray}
R_\alpha(E)q_3 &=& -iq_1\sin\sqrt{E}+q_2|T_\alpha(E)|,\label{Neu.1}\\ R_\alpha(E)q_4 &=&
-iq_2\sin\sqrt{E}+\frac{q_1}{b(\alpha)}|T_\alpha(E)|.\label{Neu.2}
\end{eqnarray}
From both (\ref{Q.off}) and (\ref{Neu.2}) it follows that $b(\alpha)^{-1}$ is real
analytic with respect to $\alpha\in\R$, and thus the relations (\ref{Neu.1}),
(\ref{Neu.2}) hold for all $\alpha\in\R$. Taking $\alpha=0$ gives
\begin{eqnarray*}
-iq_1\sin\sqrt{E}+q_2=0,\\
\frac{q_1}{b(0)}-iq_2\sin\sqrt{E}=0.
\end{eqnarray*}
The existence of a nontrivial solution in $q_1$ and $q_2$ of this system implies that
$b(0)=-1$. Thus, we obtain $q_1=-i\sin\sqrt{E}q_2$ and $q_1q_2\neq 0$. Inserting this in
(\ref{Neu.1}) we obtain $R_\alpha(E)q_3=q_2(|T_\alpha(E)|-1)$ for all $\alpha\in\R$. This
together with $|T_\alpha(E)|^2+|R_\alpha(E)|^2=1$ obviously implies that $|R_\alpha(E)|$
is constant and thus $R_\alpha(E)=0$ for all $\alpha\in\R$. Thus $E\in S$. This completes
the discussion of the case (ii).

Now consider the case (iii). As it is easy to see in this case that all
$\widetilde{\Lambda}_\alpha(E)$ have a common eigenvector, such that only one of the
functions $p(g_\alpha^{(\pm)}(E))$ is constant or
$p(g_\alpha^{(+)}(E))=p(g_\alpha^{(-)}(E))=\const$ (and thus
$\cos^2(\sqrt{E}-\pi\xi_\alpha(E))=|T_\alpha(E)|^2$) for all $\alpha\in\supp\varphi$. In
the first case either $E\in S^{(+)}$ or $E\in S^{(-)}$ but $E\notin S^{(+)}\cap S^{(-)}$.
In the second case the matrices $\widetilde{\Lambda}_{\alpha}(E)$ are not diagonalizable
and
\begin{eqnarray*}
E&\in& S^{(+)}\cap\left\{E>0:\
\cos^2(\sqrt{E}-\pi\xi_\alpha(E))=|T_\alpha(E)|^2\
\textrm{for all}\ \alpha\in\R\right\}\\
&=& S^{(-)}\cap\left\{E>0:\
\cos^2(\sqrt{E}-\pi\xi_\alpha(E))=|T_\alpha(E)|^2\
\textrm{for all}\ \alpha\in\R\right\}.
\end{eqnarray*}
Conversely, if one of $p(g_\alpha^{(\pm)}(E))$ (say $p(g_\alpha^{(+)}(E))$) does not
depend on $\alpha$, then the matrix
\begin{displaymath}
Q=\left(\begin{array}{cc} 1 & 0 \\ p(g_\alpha^{(+)}(E)) & -1\end{array} \right)
\end{displaymath}
is such that
\begin{displaymath}
Q^{-1}\widetilde{\Lambda}_\alpha(E)Q=\left(\begin{array}{cc} \lambda_+(\alpha) &
\frac{R_\alpha(E)}{T_\alpha(E)}
\\ 0 &
\lambda_-(\alpha)\end{array} \right).
\end{displaymath}

Let $E\in S^{(+)}$ and $\cos^2(\sqrt{E}-\pi\xi_\alpha(E))\leq|T_\alpha(E)|^2$ for all
$\alpha$. Then all eigenvalues of all $\widetilde{\Lambda}_\alpha(E)$ lie on the unit
circle. We have
\begin{displaymath}
\prod_{j=-n}^m\widetilde{\Lambda}_{\alpha_j(\omega)}(E)=Q\cdot\prod_{j=-n}^m
\left(\begin{array}{cc} \lambda_+(\alpha_j(\omega)) & \frac{R_{\alpha_j(\omega)}}
{T_{\alpha_j(\omega)}}\\ 0 & \lambda_-(\alpha_j(\omega))\end{array}\right)\cdot Q^{-1}.
\end{displaymath}
Obviously, there is $\delta>0$ such that
\begin{displaymath}
\frac{|R_\alpha(E)|}{|T_\alpha(E)|}\leq \delta
\end{displaymath}
for all $\alpha\in\supp\varphi$. It is easy to see that if the numbers $\beta_j$ are such
that $|\beta_j|=1$, then
\begin{equation}\label{Kreutz}
\prod_{j=-n}^m\left(\begin{array}{cc} \beta_j & b_j \\
                                      0 & \overline{\beta_j} \end{array}\right)=
\left(\begin{array}{cc}\prod_{j=-n}^m\beta_j & b\\
                        0 & \prod_{j=-n}^m\overline{\beta_j}\end{array}\right)
\end{equation}
with $b$ satisfying the inequality $|b|\leq\sum_{j=-n}^m|b_j|$. Thus, the norm of the
matrix on the r.h.s. of (\ref{Kreutz}) is less or equal $\sqrt{2+|b|^2}$. Therefore,
\begin{displaymath}
\left\|\prod_{j=-n}^m\widetilde{\Lambda}_{\alpha_j(\omega)}(E)\right\|\leq
\left[2+(n+m+1)^2\delta^2 \right]^{1/2},
\end{displaymath}
and thus $\gamma(E)=0$.

Let now $E\in S^{(+)}$ and $\cos^2(\sqrt{E}-\pi\xi_\alpha(E))>|T_\alpha(E)|^2$ for almost
$\alpha\in\supp\varphi$. In this case either $|\lambda_+(\alpha)|>1$ for almost all
$\alpha\in\supp\varphi$ or $|\lambda_+(\alpha)|<1$ for almost all $\alpha\in\supp\varphi$.
The eigenvalue of $\prod_{j=-n}^m\widetilde{\Lambda}_{\alpha_j(\omega)}(E)$ corresponding
to $g_\alpha^{(+)}(E)$ is given by $\prod_{j=-n}^m\lambda_+(\alpha_j(\omega))$. Therefore
\begin{displaymath}
\log\left\|\prod_{j=-n}^m\widetilde{\Lambda}_{\alpha_j(\omega)}(E) \right\|
\geq \sum_{j=-n}^m\log|\lambda_+(\alpha_j(\omega))|
\end{displaymath}
if $|\lambda_+(\alpha)|>1$, or
\begin{displaymath}
\log\left\|\prod_{j=-n}^m\widetilde{\Lambda}_{\alpha_j(\omega)}(E) \right\|
\geq -\sum_{j=-n}^m\log|\lambda_+(\alpha_j(\omega))|
\end{displaymath}
if $|\lambda_+(\alpha)|<1$. By the Birkhoff-Khintchin theorem we have
\begin{displaymath}
\lim_{m,n\rightarrow\infty}\frac{1}{m+n+1}\sum_{j=-n}^m\log|\lambda_+(\alpha_j(\omega))|=
\E\left\{\log|\lambda_+(\alpha_j(\omega))| \right\}
\end{displaymath}
almost surely. If $|\lambda_+(\alpha)|>1$ ($<1$) for almost all $\alpha$, we
have $\E\left\{\log|\lambda_+(\alpha_j(\omega))| \right\}>0$ ($<0$), and hence
\begin{displaymath}
\gamma(E)=\lim_{m,n\rightarrow\infty}\frac{1}{m+n+1}
\log\left\|\prod_{j=-n}^m\widetilde{\Lambda}_{\alpha_j(\omega)}(E)\right\|>0
\end{displaymath}
almost surely. Similarly, in the case $E\in S^{(-)}$ and
$\cos^2(\sqrt{E}-\pi\xi_\alpha(E))>|T_\alpha(E)|^2$ for almost all $\alpha\in\supp\varphi$
we have $\gamma(E)>0$ almost surely.

Thus, we have shown that the cases (i), (ii), (iii) occur iff $E\in S\cup S^{(+)}\cup
S^{(-)}$. Moreover, $\gamma(E)=0$ for $E\in\widetilde{S}$ and $\gamma(E)>0$ almost surely
for $E\in(S^{(+)}\cup S^{(-)})\setminus \widetilde{S}$. For those and only those $E>0$,
which do not belong to $S\cup S^{(+)}\cup S^{(-)}$ the case (iv) occurs, and thus the
second condition of the Furstenberg theorem is fulfilled.

We turn to the first condition of the Furstenberg theorem.

If $E$ is in a gap of $H^{(\alpha)}$ for at least one $\alpha\in\supp \varphi$, then one
of the eigenvalues (\ref{eigenvalues}) is strictly larger than 1 in absolute value.
Therefore $\|\widetilde{\Lambda}_\alpha (E)^n\|\geq|\max_\pm\lambda_\pm(\alpha)|^n
\rightarrow\infty$ as $n\rightarrow\infty$, and hence $G_E$ is not compact.

Now suppose that $E$ is not in a gap of $H^{(\alpha)}$ for all $\alpha\in\supp\varphi$,
i.e. $\cos^2(\sqrt{E}-\pi\xi_\alpha(E))\leq |T_\alpha(E)|^2$ for such $\alpha$. In this
case the eigenvalues $\lambda_\pm(\alpha)$ lie on the unit circle.

Since $\supp\varphi$ contains at least two $\alpha$'s such that the corresponding
$\widetilde{\Lambda}_\alpha$'s have no common eigenvectors, by the assumed continuity of
$\varphi$ we can select them in such a way that either
$\cos^2(\sqrt{E}-\pi\xi_\alpha(E))<|T_\alpha(E)|^2$ or
$\cos^2(\sqrt{E}-\pi\xi_\alpha(E))=|T_\alpha(E)|^2$ for both $\alpha$'s. In the first case
we can apply the arguments of \cite{Matsuda:Ishii,Ishii} to construct a matrix, which
belongs to $G_E$ and has an eigenvalue strictly larger than 1. To treat the second case we
consider two matrices
\begin{displaymath}
M_j=\left(\begin{array}{cc}
           a_j & b_j \\
           b_j^\ast & a_j^\ast
           \end{array}\right)
\end{displaymath}
with no common eigenvectors, satisfying $|{\Re}a_j|=1$ for both $j=1,2$. Since
$M_j\in\SL(2;\C)$ one has $|{\Im}a_j|=|b_j|$. The matrices $M_j$ are not diagonalizable,
because to the eigenvalue $\lambda_j={\Re}a_j$ corresponds the only eigenvector
$g_j=(ib_j, {\Im}a_j)^T$. Further we consider the matrices $\widetilde{M}_1$,
$\widetilde{M}_2$, which are defined as follows: $\widetilde{M}_j=M_j^2$, $j=1,2$ if
${\Re}a_1{\Im}a_1{\Re}a_2{\Im}a_2<0$; $\widetilde{M}_1=M_1^2$, $\widetilde{M}_2=M_2^{-2}$
if ${\Re}a_1{\Im}a_1{\Re}a_2{\Im}a_2>0$. These matrices have the form
\begin{displaymath}
\widetilde{M}_j=\left(\begin{array}{cc}
                 \widetilde{a}_j & \widetilde{b}_j \\
                 \widetilde{b}^\ast_j & \widetilde{a}^\ast_j
                 \end{array}\right),
\end{displaymath}
such that ${\Re}\widetilde{a}_j=({\Re}a_j)^2=1$ and
${\Im}\widetilde{a}_1{\Im}\widetilde{a}_2<0$. We can write the matrices $\widetilde{M}_j$
in the following form
\begin{displaymath}
\widetilde{M}_j=I+i{\Im}\widetilde{a}_jN_j,
\end{displaymath}
with
\begin{displaymath}
N_j=\left(\begin{array}{cc}
          1 & e^{i\vartheta_j} \\
          -e^{-i\vartheta_j} & -1
          \end{array}\right),
\end{displaymath}
where $e^{i\vartheta_j}=-i\widetilde{b}_j/{\Im}\widetilde{a}_j=-ib_j/{\Im}a_j$. Obviously,
$N_j^2=0$. Since by assumption the eigenvectors of the matrices $M_j$ are linearly
independent, we have that $\vartheta_1\neq\vartheta_2$. Now we calculate
\begin{displaymath}
\tr(\widetilde{M}_1\widetilde{M}_2)=2-4{\Im}\widetilde{a}_1{\Im}\widetilde{a}_2
\sin^2\frac{\vartheta_1-\vartheta_2}{2}>2.
\end{displaymath}
Therefore, one of the eigenvalues of $\widetilde{M}_1\widetilde{M}_2$ is strictly larger
than 1.

Thus, by the Furstenberg theorem it follows that $\gamma(E)>0$ almost surely for $E\notin
S\cup S^{(+)}\cup S^{(-)}$. This completes the proof of the theorem. $\blacksquare$

\vspace{0.25in}

Let $(\cdot,\cdot)$ denote the inner product in $\C^2$. We note that
\begin{displaymath}
\xi^{(-n,m)}(E;\omega)=\pm\frac{1}{\pi}\arg\left(
e_\pm, \Lambda^{(-n,m)}(E;\omega)e_\pm \right)
\end{displaymath}
with $e_+=(1,0)^T$, $e_-=(0,1)^T$. From (\ref{Lambda.U}) it follows that
\begin{eqnarray}\label{Stern.2}
\left(e_\pm, \Lambda^{(-n,m)}(E;\omega)e_\pm \right)=
\left(e_\pm, U_E^{-n-1/2}\prod_{j=-n}^m\widetilde{\Lambda}_{\alpha_j(\omega)}(E)
U_E^{-m-1/2}e_\pm\right)\nonumber\\
=\left(U_E^{n+1/2}e_\pm, \prod_{j=-n}^m\widetilde{\Lambda}_{\alpha_j(\omega)}(E)
U_E^{-m-1/2}e_\pm \right).
\end{eqnarray}
Obviously, $U_E e_\pm=e^{\mp i\sqrt{E}}e_\pm$. Therefore the r.h.s. of (\ref{Stern.2})
equals
\begin{displaymath}
e^{\pm i\sqrt{E}(n+m+1)}\left(e_\pm, \prod_{j=-n}^m
\widetilde{\Lambda}_{\alpha_j(\omega)}(E) e_\pm\right).
\end{displaymath}
Thus we obtain
\begin{displaymath}
\xi(E)=\frac{\sqrt{E}}{\pi}\pm\frac{1}{\pi}\lim_{n,m\rightarrow\infty}\frac{1}{n+m+1}
\arg\left(e_\pm, \prod_{j=-n}^m \widetilde{\Lambda}_{\alpha_j(\omega)}(E) e_\pm\right).
\end{displaymath}
Since $\xi(E)=N_0(E)-N(E)=\sqrt{E}/\pi-N(E)$, it follows that
\begin{displaymath}
N(E)=\mp\frac{1}{\pi}\lim_{n,m\rightarrow\infty}\frac{1}{n+m+1}
\arg\left(e_\pm, \prod_{j=-n}^m \widetilde{\Lambda}_{\alpha_j(\omega)}(E) e_\pm\right).
\end{displaymath}
This representation is similar to the definition of the density of states through the
rotation number of fundamental solution of the Schr\"odinger equation
\cite{Johnson:Moser}.

The representations (\ref{gamma.E.1}), (\ref{gamma.E.2}) can also be rewritten in a
similar form,
\begin{eqnarray*}
\gamma(E) &=& \lim_{m,n\rightarrow\infty}\frac{1}{m+n+1}\log\left|\left(e_\pm,
\Lambda^{(-n,m)}(E) \right)\right|\\
&=& \lim_{m,n\rightarrow\infty}\frac{1}{m+n+1}\log\left|\left(e_\pm,
\prod_{j=-n}^m\widetilde{\Lambda}_{\alpha_j(\omega)}(E)e_\pm \right)\right|.
\end{eqnarray*}

\vspace{0.25in}

\textbf{Example 1.} Here we consider the Hamiltonian (\ref{1.1}) where $f$ is (formally)
replaced by the Dirac $\delta$-function. In this case the transmission and reflection
amplitudes are given by
\begin{eqnarray*}
T_\alpha(E) &=&
\left(1+\frac{i\alpha}{2\sqrt{E}}\right)^{-1},\\
R_\alpha(E) &=& -i\frac{\alpha}{2\sqrt{E}}\left(1+\frac{i\alpha}{2\sqrt{E}}
\right)^{-1}.
\end{eqnarray*}
Therefore $S=\emptyset$ and
\begin{eqnarray*}
F_E^{(\pm)}(\alpha)&=& \left(\frac{2\sqrt{E}}{\alpha}\sin\sqrt{E}+\cos\sqrt{E}
\right)\\
&&\pm i\frac{2\sqrt{E}}{\alpha}\sqrt{\left(\cos\sqrt{E}+\frac{\alpha}
{2\sqrt{E}}\sin\sqrt{E}\right)^2-1},\\
\frac{\cos^2(\sqrt{E}-\pi\xi_\alpha(E))}{|T_\alpha(E)|^2} &=& \left(\cos\sqrt{E}+\frac{\alpha}
{2\sqrt{E}}\sin\sqrt{E}\right)^2.
\end{eqnarray*}
Thus $\widetilde{S}=\{(\pi k)^2,\ k\in\N\}$. From Theorem 5.6 it follows that $\gamma(E)$
is positive on $(0,\infty)$ except for the set $\widetilde{S}$. That $\gamma(E)=0$ iff
$E=E_k=(\pi k)^2$, $k\in\Z$ was proved by Ishii \cite{Ishii} (see also
\cite{Figotin:Pastur}). Kirsch and Nitzschner \cite{Kirsch:Nitzschner} proved that
$N(E_k)=N_0(E_k)$. Here we reconsider these facts once more.

The matrices $\Lambda_j(E)$ can be calculated explicitly:
\begin{displaymath}
\Lambda_j(E)=I+\frac{i\alpha_j(\omega)}{2\sqrt{E}}A_j(E),
\end{displaymath}
where
\begin{displaymath}
A_j(E)=\left(\begin{array}{lr}
             1 & e^{-2i\sqrt{E}j} \\
             -e^{2i\sqrt{E}j} & -1
             \end{array}\right).
\end{displaymath}
For $E=E_k$, $k\in\N$
\begin{displaymath}
A_j(E_k)=A=\left(\begin{array}{lr}
             1 & 1 \\
             -1 & -1
             \end{array}\right).
\end{displaymath}
Obviously $A$ is nilpotent, i.e. $A^2=0$. Therefore
\begin{eqnarray*}
\Lambda^{(-n,m)}(E_k;\omega)=\prod_{j=-n}^m\Lambda_j(E_k)=\prod_{j=-n}^m\left(I+
\frac{i\alpha_j(\omega)}{2\sqrt{E_k}}A \right)\\
=I+\frac{i}{2\sqrt{E_k}}A\sum_{j=-n}^m \alpha_j(\omega).
\end{eqnarray*}
From this it follows that
\begin{displaymath}
\left(e_\pm, \Lambda^{(-n,m)}(E_k;\omega)e_\pm\right)=1\pm\frac{i}{2\sqrt{E_k}}
\sum_{j=-n}^m\alpha_j(\omega),
\end{displaymath}
and hence
\begin{eqnarray*}
\arg\left(e_\pm, \Lambda^{(-n,m)}(E_k;\omega)e_\pm\right) &=&
\pm\arctan\left(\frac{1}{2\sqrt{E_k}}
\sum_{j=-n}^n \alpha_j(\omega) \right),\\
\left|\left(e_\pm, \Lambda^{(-n,m)}(E_k;\omega)e_\pm\right) \right|^2 &=&
1+\frac{1}{4E_k}
\left( \sum_{j=-n}^n \alpha_j(\omega)\right)^2 .
\end{eqnarray*}
Since $\{\alpha_j(\omega) \}_{j\in\Z}$ is metrically transitive, by the Birkhoff --
Khintchin ergodic theorem we have
\begin{eqnarray*}
\lim_{n,m\rightarrow\infty}\frac{1}{m+n+1}\sum_{j=-n}^m \alpha_j(\omega)
=\lim_{n,m\rightarrow\infty}\frac{1}{m+n+1}\sum_{j=-n}^m \alpha_0(T_j\omega)
=\E\{\alpha\}.
\end{eqnarray*}
Therefore, the sum $\sum_{j=-n}^m\alpha_j(\omega)$ in the limit $m,n\rightarrow\infty$
increases not faster than $(n+m+1)$. Hence
\begin{eqnarray*}
\xi(E_k)=\frac{1}{\pi}\lim_{m,n\rightarrow\infty}\frac{1}{m+n+1}\arg
\left(e_\pm, \Lambda^{(-n,m)}(E_k;\omega)e_\pm\right)=0,\\
\gamma(E_k)=\lim_{m,n\rightarrow\infty}\frac{1}{m+n+1}\log
\left|\left(e_\pm, \Lambda^{(-n,m)}(E_k;\omega)e_\pm\right) \right|=0.
\end{eqnarray*}

\section{Analytic Continuation and the Thouless Formula}
\setcounter{equation}{0}
\markright{6 Analytic Continuation and the Thouless Formula}

In this Section we show that the analyticity of $w(E)=-\gamma(E)+i\pi N(E)$ in the upper
complex $E$-plane $\C_+$ and the Thouless formula (\ref{Thouless}) are a direct
consequence of the fact that the functions $\log
T^{(-n,m)}(E)=\log|T^{(-n,m)}(E)|+i\pi\xi^{(-n,m)}(E)$ are analytic in $\C_+$ for every
$n,m\in\N$.

By Lemma 2.3 we have
\begin{displaymath}
\log T^{(-n,m)}(z)=-\int_\R\frac{\xi_\omega^{(-n,m)}dE}{E-z}.
\end{displaymath}

\vspace{0.25in}

\textbf{Lemma 6.1} \it Let $E_0=\max_\pm\left\{|\alpha_\pm|\right\}f_0$ with $f_0=\sup|f(x)|$.
Then there is a constant $C$ independent of $E$ and $n,m\in\N$ such that
\begin{displaymath}
|\xi_\omega^{(-n,m)}(E)|\leq \frac{C}{\sqrt{E}}(n+m+1)
\end{displaymath}
for all $E>E_0$. \rm

\vspace{0.25in}

\textit{Proof.} By the monotonicity of the spectral shift function we have
\begin{displaymath}
|\xi_\omega^{(-n,m)}(E)|\leq\xi(E;H_0+E_0\chi^{(-n,m)},H_0),
\end{displaymath}
where $\chi^{(-n,m)}$ is the characteristic function of the interval $[-n-1/2,m+1/2]$.
Calculating $\xi(E;H_0+E_0\chi^{(-n,m)},H_0)$ explicitly we find
\begin{eqnarray*}
\lefteqn{\xi(E;H_0+E_0\chi^{(-n,m)},H_0)
=\frac{\sqrt{E}}{\pi}(n+m+1)}\\&-&
\frac{1}{2\pi i}\log\frac{(\sqrt{E-E_0}+\sqrt{E})^2e^{i\sqrt{E-E_0}(n+m+1)}-
(\sqrt{E-E_0}-\sqrt{E})^2e^{-i\sqrt{E-E_0}(n+m+1)}}
{(\sqrt{E-E_0}+\sqrt{E})^2e^{-i\sqrt{E-E_0}(n+m+1)}-
(\sqrt{E-E_0}-\sqrt{E})^2e^{i\sqrt{E-E_0}(n+m+1)}}\\ &=& \frac{\sqrt{E}}{\pi}(n+m+1)-
\frac{1}{\pi}\mathrm{Arctan}\left(\frac{2E-E_0}{2\sqrt{E}\sqrt{E-E_0}}
\tan(\sqrt{E-E_0}(n+m+1)) \right),
\end{eqnarray*}
where $\mathrm{Arctan}$ is the multivalued $\arctan$ function such that
$\mathrm{Arctan}(C\tan(x))$ is continuous and nondecreasing with respect to $x$. Since
\begin{displaymath}
\frac{2E-E_0}{2\sqrt{E}\sqrt{E-E_0}}\geq 1
\end{displaymath}
for all $E>E_0$, it follows that
\begin{displaymath}
\mathrm{Arctan}\left(\frac{2E-E_0}{2\sqrt{E}\sqrt{E-E_0}}
\tan x \right)\geq\mathrm{Arctan}(\tan x)=x.
\end{displaymath}
Therefore
\begin{displaymath}
\xi(E;H_0+E_0\chi^{(-n,m)},H_0)\leq \frac{\sqrt{E}-\sqrt{E-E_0}}{\pi}(n+m+1).
\end{displaymath}
Obviously,
\begin{displaymath}
\sqrt{E}-\sqrt{E-E_0}\leq\frac{E_0}{\sqrt{E}}
\end{displaymath}
for all $E\geq E_0$, thus proving the lemma. $\blacksquare$

\vspace{0.25in}

Now to study the limit $n,m\rightarrow\infty$ we can use the theorem on the continuity of
the Stieltjes transform (see e.g. \cite[Appendix A]{Figotin:Pastur}). The applicability of
this theorem is guaranteed by Lemma 6.1, from which it follows that
\begin{displaymath}
\int_\R\frac{|\xi_\omega^{(-n,m)}(E)|}{1+|E|}dE<\infty,
\end{displaymath}
and
\begin{displaymath}
\lim_{c\rightarrow\infty}\sup_{n,m}\ (n+m+1)^{-1}
\int_c^\infty\frac{|\xi_\omega^{(-n,m)}(E)|}{|E|}dE = 0.
\end{displaymath}
Therefore, since $(n+m+1)^{-1}\xi_\omega^{(-n,m)}(E)\rightarrow \xi(E)$ for all $E\in\R$
and $\P$-almost all $\omega\in\Omega$, and $\xi(E)$ is nonrandom, we obtain that
$(n+m+1)^{-1}\log T_\omega^{(-n,m)}(E)$ converges for all $z\in\C$ with ${\Im} z>0$ and
$\P$-almost all $\omega\in\Omega$ to some deterministic limit $W(z)$, which is given by
\begin{equation}\label{integ.repr}
W(z)=-\int_\R\frac{\xi(E)dE}{E-z},
\end{equation}
and therefore is analytic for ${\Im} z>0$.

Since $\xi(E)$ is continuous (moreover H\"older continuous) by the Sokhotski-Plemelj
formula we have that $W(E+i0)$ exists for all $E\in\R$ and
\begin{displaymath}
{\Im} W(E+i0)=i\pi\xi(E).
\end{displaymath}
As in the proof of Lemma 2.3, since $\xi(E)$ is continuous and is of bounded variation, we
get
\begin{equation}\label{ReW}
{\Re} W(E+i0)=\int_\R \log|\lambda-E|d\xi(\lambda).
\end{equation}
On the other hand, by Lemma 2.3 we have
\begin{equation}\label{logT}
\log|T_\omega^{(-n,m)}(E)|=\int_\R\log|\lambda-E|d\xi^{(-n,m)}(\lambda;\omega).
\end{equation}
By Theorem 5.1 we have that for every fixed $E>0$
\begin{equation}\label{gamE}
-\gamma(E)=\lim_{n,m\rightarrow\infty}\int_\R\log|\lambda-E|\frac{d\xi^{(-n,m)}(\lambda;\omega)}
{n+m+1}
\end{equation}
almost surely. Now we prove that for Lebesgue almost all $E\in\R$ there are subsequences
$n_j$, $m_j$ tending to infinity and such that for $\P$-almost all $\omega$
\begin{equation}\label{6.5}
\lim_{j\rightarrow\infty}\int_\R \log|\lambda-E|\frac{d\xi^{(-n_j,m_j)}(\lambda;\omega)}
{n_j+m_j+1}=\int_\R \log|\lambda-E|d\xi(\lambda).
\end{equation}
Thus, from (\ref{ReW}) and (\ref{gamE}) it will follow that ${\Re} W(E+i0)=-\gamma(E)$ and
therefore
\begin{displaymath}
\gamma(E)=-\int_\R\log|\lambda-E|d\xi(\lambda)
\end{displaymath}
for almost all $E>0$.

The arguments used below are very similar to those of Pastur and Figotin \cite[Theorem
11.6]{Figotin:Pastur}. Consider the functions
\begin{eqnarray*}
\tau(E) &=& \int_\R\log|\lambda-E|d\xi(\lambda)=
\int_\R\log\left|
\frac{\lambda-E}{i-E} \right|d\xi(\lambda),\\
\tau_\omega^{(-n,m)}(E) &=& \int_\R\log|\lambda-E|\frac{d\xi^{(-n,m)}(\lambda)}{n+m+1}=
\int_\R\log\left|
\frac{\lambda-E}{i-E} \right|\frac{d\xi^{(-n,m)}(\lambda)}{n+m+1}.
\end{eqnarray*}
Let us fix some interval $K\subset\R$ and consider
\begin{displaymath}
t_B(\lambda)=\int_B \log\left|\frac{\lambda-E}{i-E} \right|dE,
\end{displaymath}
where $B\in\mathcal{B}(K)$ (the set of all Borel subsets in $K$). The family of functions
$\{t_B(\lambda), B\in\mathcal{B}(K) \}$ is uniformly bounded and equicontinuous on any
bounded interval and
\begin{displaymath}
\sup_{B\in\mathcal{B}(K)}|t_B(\lambda)|\leq C(1+|\lambda|)^{-1}
\end{displaymath}
for some $C>0$. Also we have
\begin{displaymath}
\int_B\tau(E)dE=\int_\R t_B(\lambda)d\xi(\lambda).
\end{displaymath}
Since the family $t_B(\lambda)$ is uniformly bounded and equicontinuous and since
$(n+m+1)^{-1}d\xi^{(-n,m)}(\lambda;\omega)$ converges vaguely to $d\xi(\lambda)$ almost
surely, it follows that
\begin{equation}\label{Stern}
\lim_{n,m\rightarrow\infty}\sup_{B\in\mathcal{B}(K)}\left|
\int_B \left(\tau_\omega^{(-n,m)}(E)-\tau(E) \right)dE \right|=0.
\end{equation}
On the other hand one has
\begin{eqnarray*}
&& \int_K|\tau_\omega^{(-n,m)}(E)-\tau(E)|dE \\ &=&
\int_{K\cap\{\tau_\omega^{(-n,m)}(E)\geq\tau(E) \}}
\left(\tau_\omega^{(-n,m)}(E)-\tau(E) \right)dE\\
&&-
\int_{K\cap\{\tau_\omega^{(-n,m)}(E)<\tau(E) \}}
\left(\tau_\omega^{(-n,m)}(E)-\tau(E) \right)dE\\
&\leq& 2\sup_{B\in\mathcal{B}(K)}\left|
\int_B \left(\tau_\omega^{(-n,m)}(E)-\tau(E) \right)dE \right|.
\end{eqnarray*}
From (\ref{Stern}) it follows that
\begin{displaymath}
\lim_{n,m\rightarrow\infty}\int_K|\tau_\omega^{(-n,m)}(E)-\tau(E)|dE=0.
\end{displaymath}

Denote by $\Omega_1$ the subset of $\Omega$ such that $\P(\Omega_1)=1$ and for which
$(n+m+1)^{-1}$ $d\xi^{(-n,m)}(\lambda;\omega)$ converges to $d\xi(\lambda)$. Taking an
arbitrary $\omega\in\Omega_1$ we can choose subsequences $n_j$, $m_j$ tending to infinity
and such that
\begin{equation}\label{Stern2}
\lim_{j\rightarrow\infty}|\tau_\omega^{(-n_j,m_j)}(E)-\tau(E)|=0
\end{equation}
for Lebesgue almost all $E\in K$. Since
$(n+m+1)^{-1}d\xi^{(-n,m)}(\lambda;\omega)\rightarrow d\xi(\lambda)$ for almost all
$\omega\in\Omega$, relation (\ref{Stern2}) remains valid for almost every
$\omega\in\Omega$. Thus (\ref{6.5}) is proven.

We have shown that
\begin{eqnarray*}
{\Im} W(E+i0)=i\pi\xi(E)\\ {\Re} W(E+i0)=-\gamma(E)
\end{eqnarray*}
for almost all $E>0$. Therefore, $W(z)$ is an analytic continuation of
$-\gamma(E)+i\pi\xi(E)$ from the real semiaxis.

Let us consider the function $w(z)=W(z)+w_0(z)$, where
\begin{displaymath}
w_0(z)=-\gamma_0(z)+i\pi N_0(z)=-\sqrt{-z}
\end{displaymath}
with $\gamma_0(E)$ being the Lyapunov exponent for $H_0=-d^2/dx^2$. It is obviously
analytic in the upper half-plane ${\Im} E> 0$ and
\begin{displaymath}
w(E+i0) = -\gamma(E)+i\pi N(E),\ E>0.
\end{displaymath}

From (\ref{integ.repr}) and from the fact that $\xi(E)$ is given by the difference of two
nonnegative functions $N_0(E)$ and $N(E)$, it follows that $W(z)$ is the difference of two
Nevanlinna functions. Since $w_0(z)$ is a Nevanlinna function, so is $w(z)$. Also from
(\ref{6.5}) it follows that the Thouless formula (\ref{Thouless}) holds.

\section{Some Extensions}
\setcounter{equation}{0}
\markright{7 Some Extensions}

Let $V(x)$ be some real-valued uniformly locally integrable function. For simplicity we
suppose that $V(x)$ is uniformly bounded, i.e. $|V(x)|\leq V_0$, $V_0>0$. The last
assumption can be weakend but we do not go into details here. Let $\{y_j\}_{j\in\Z}$ be a
sequence of real numbers such that $y_j\rightarrow\pm\infty$ as $j\rightarrow\pm\infty$
and $y_j< y_{j+1}$ for all $j\in\Z$. We suppose that $\{y_j\}_{j\in\Z}$ is such that all
the differences $y_{j+1}-y_j$ are finite (but not necessarily uniformly bounded). Let
$\chi_j(x)$ be the characteristic function of the interval $[y_j,y_{j+1}]$. We denote
$V_j(x)=V(x)\chi_j(x)$ such that
\begin{displaymath}
V^{(-n,m)}(x)=\sum_{j=-n}^m V_j(x)
\end{displaymath}
tends to $V(x)$ as $m,n\rightarrow\infty$. By the above assumption one has $|V_j(x)|\leq
V_0\chi_j(x)$.

Let $H$, $H_j$ and $H^{(-n,m)}$ denote the Hamiltonians with domains of definition being
the Sobolev space $W^{(2,2)}(\R)$,
\begin{displaymath}
H=H_0+V,\ H_j=H_0+V_j,\ H^{(-n,m)}=H_0+V^{(-n,m)}.
\end{displaymath}
Let $\xi_j(E)$, $T_j(E)$ and $\xi^{(-n,m)}(E)$, $T^{(-n,m)}(E)$ be the spectral shift
function and transmission amplitude for the pairs ($H_j$, $H_0$) and ($H^{(-n,m)}$,
$H_0$), respectively. Also as above we denote
\begin{displaymath}
\widetilde{\xi}_\pm^{(-n,m)}(E)=\xi^{(-n,m)}(E)\pm 1.
\end{displaymath}
By Corollary 3.4 and Theorem 5.2 we have
\begin{eqnarray*}
\widetilde{\xi}_+^{(-n,m)}(E) &\leq& \widetilde{\xi}_+^{(-n,k)}(E)+
\widetilde{\xi}_+^{(k,m)}(E),\\
\widetilde{\xi}_-^{(-n,m)}(E) &\geq& \widetilde{\xi}_-^{(-n,k)}(E)+
\widetilde{\xi}_-^{(k,m)}(E),\\
|T^{(-n,m)}(E)| &\geq& \frac{1}{2}|T^{(-n,k)}(E)||T^{(k,m)}(E)|
\end{eqnarray*}
for every $k\in\Z$ such that $-n\leq k\leq m$. By the monotonicity and superadditivity
properties of the spectral shift function (Corollary 3.4) we also have
\begin{eqnarray*}
(y_m-y_{-n})^{-1}\widetilde{\xi}^{(-n,m)}_+(E) &\geq& (y_m-y_{-n})^{-1}
\left[\xi(E;H_0-V_0\chi_{[y_{-n},y_m]},H_0)+1\right]\\
&\geq& \inf_{m,n}(y_m-y_{-n})^{-1}\left[\xi(E;H_0-V_0\chi_{[y_{-n},y_m]},H_0)+1\right]\\
&=& \lim_{m,n\rightarrow\infty}
(y_m-y_{-n})^{-1}\left[\xi(E;H_0-V_0\chi_{[y_{-n},y_m]},H_0)+1\right]\\ &=&
-[\max(0,E+V_0)]^{1/2}/\pi
\end{eqnarray*}
for all $E\in\R$. Similarly,
\begin{eqnarray*}
(y_m-y_{-n})^{-1}\widetilde{\xi}^{(-n,m)}_-(E)&\leq& (y_m-y_{-n})^{-1}
\left[\xi(E;H_0+V_0\chi_{[y_{-n},y_m]},H_0)-1\right]\\
&\leq& [\max(0,E-V_0)]^{1/2}/\pi.
\end{eqnarray*}
Also we have that $|T^{(-n,m)}(E)|<1$. Therefore, if $H_0+V^{(-n,m)}$ and
$H_0+V^{(-n+k,m+k)}$ are unitary equivalent from the known property of
subadditive functions (see e.g. \cite[Theorem 6.6.1]{Hille}) the existence of
the limits
\begin{eqnarray*}
\xi(E) &=& \lim_{m,n\rightarrow\infty}\frac{\xi^{(-n,m)}(E)}{y_m-y_{-n}},\\
\gamma^+_T(E) &=& -\lim_{m\rightarrow\infty}\frac{\log|T^{(0,m)}(E)|}{y_m},\\
\gamma^-_T(E) &=& -\lim_{n\rightarrow\infty}\frac{\log|T^{(-n,0)}(E)|}{|y_{-n}|},\\
\gamma_T(E) &=& -\lim_{m,n\rightarrow\infty}\frac{\log|T^{(-n,m)}(E)|}{y_m-y_{-n}}
\end{eqnarray*}
follows. Clearly $\gamma_T^+$, $\gamma_T^-$, and $\gamma_T$ may be unequal. Theorems 4.3
and 5.3 apply also to this case. Thus we have again that $\xi(E)=N_0(E)-N(E)$ for all
$E\in\R$ and $\gamma^\pm_T(E)=\gamma^\pm(E)$, $\gamma_T(E)=\gamma(E)$ for all positive
$E$, where
\begin{eqnarray*}
\gamma^\pm(E) &=& \lim_{x\rightarrow\pm\infty}\frac{1}{|x|}\log\|\phi(0,x;E)\|,\\
\gamma(E) &=& \lim_{x\rightarrow\infty}\frac{1}{2x}\log\|\phi(-x,x;E)\|
\end{eqnarray*}
are the upper Lyapunov exponents. Here $\phi(x,x';E)$ denotes the fundamental matrix of
the Schr\"odinger equation with the potential $V$. Further we can again prove that
\begin{eqnarray*}
\gamma(E)&=&\lim_{m,n\rightarrow\infty}\frac{1}{y_m-y_{-n}}\log\left\|\prod_{j=-n}^m
\widetilde{\Lambda}_j(E)\right\|,\\
N(E)&=& \mp\frac{1}{\pi}\lim_{m,n\rightarrow\infty}\frac{1}{y_m-y_{-n}}
\arg\left(e_\pm,\prod_{j=-n}^m\widetilde{\Lambda}_j(E)e_\pm\right),
\end{eqnarray*}
where
\begin{displaymath}
\widetilde{\Lambda}_j(E)=\left(\begin{array}{cc}
                               \frac{e^{i\sqrt{E}(y_{j+1}-y_{j-1})/2}}{T_j(E)} &
                               -\frac{R_j(E)}{T_j(E)}e^{-i\sqrt{E}(y_{j+1}-2y_j+y_{j-1})/2}\\
                               \frac{L_j(E)}{T_j(E)}e^{i\sqrt{E}(y_{j+1}-2y_j+y_{j-1})/2} &
                                \frac{e^{-i\sqrt{E}(y_{j+1}-y_{j-1})/2}}{T_j(E)^\ast}
                               \end{array}\right),
\end{displaymath}
and $T_j(E)$, $R_j(E)$, $L_j(E)$ are transmission and reflection amplitudes corresponding
to the potential $V_j(\cdot-y_j)$.

\section*{Appendix A}
\renewcommand{\theequation}{A.\arabic{equation}}
\setcounter{equation}{0}
\markright{Appendix A}

Here we prove that the expression in the braces on the r.h.s. of (\ref{long.1}) is bounded
from below by the constant $C(E)=4E/(1+E^2)$. For brevity we set
$B=\sqrt{1-|T_\omega^{1,n}|^2}$, $a=\sin\theta_1$, and $b=\sin\theta_2$. Then the
expression at hand can be written as follows
\begin{eqnarray*}
F(a,b,E,B) &=& 2B^2(1-a^2)+2(1-b^2)\nonumber\\ && +\frac{1}{E}(Ba-b)^2+E(Ba+b)^2.
\end{eqnarray*}
We show that $F(a,b)\geq C(E)$ for all $E>0$ independently of $a,b\in[-1,1]$ and
$B\in[0,1]$.

First we note that
\begin{eqnarray*}
F(a,b,E,B) &=& F(-a,-b,E,B)\nonumber\\
=F(a,-b,1/E,B) &=& F(-a,b,1/E,B).
\end{eqnarray*}
Since $C(E)=C(1/E)$ it therefore suffices to consider the case $a,b\in[0,1]$. Since
\begin{displaymath}
\frac{\partial}{\partial B}F(a,b,E,B)=2B(2-2a^2+a^2/E+a^2E)+2ab(E-1/E)
\end{displaymath}
is nonnegative whenever $E\geq 1$ we have
\begin{displaymath}
F(a,b,E,B)\geq F(a,b;E,0)=2(1-b^2)+b^2/E+E b^2\geq 2\geq C(E)
\end{displaymath}
for such $E$. Here and in what follows we use the estimate $E+1/E\geq 2$ for all $E>0$.
Now take the case $0<E<1$. Then for fixed $E$, $a$ and $b$ the function
\begin{displaymath}
\frac{\partial}{\partial B}F(a,b,E,B)
\end{displaymath}
has exactly one zero as a function of $B$ in the interval $[0,\infty)$ at
\begin{displaymath}
0<B_0=\frac{-ab(E-1/E)}{(E+1/E-2)a^2+2},
\end{displaymath}
which is a minimum for $F(a,b,E,\cdot)$. Hence
\begin{eqnarray*}
\lefteqn{F(a,b,E,B)\geq F(a,b,E,B_0)}\\
&=& -\frac{a^2b^2(E-1/E)^2}{(E+1/E-2)a^2+2}+2+(1/E+E-2)b^2\\ &=:& G(a,b,E).
\end{eqnarray*}
It is easy to see that
\begin{eqnarray*}
\frac{\partial}{\partial a}G(a,b,E)=-\frac{4ab^2(E-1/E)^2}{[(E+1/E-2)a^2+2]^2}\leq 0.
\end{eqnarray*}
Therefore
\begin{eqnarray*}
G(a,b,E) &\geq& G(1,b,E) = 2+ 2 b^2\frac{2-1/E-E}{E+1/E}\\ &\geq &
G(1,1,E)=2+2\frac{2-1/E-E}{E+1/E}=\frac{4E}{1+E^2},
\end{eqnarray*}
thus proving the claim.

\newpage

\markright{References}

\end{document}